%% file: main.tex
\documentclass[twocolumn, astrosymb]{aastex631} % , linenumbers

\usepackage{graphicx}
\usepackage{amsmath}
\usepackage{amssymb}
\usepackage{makecell}
\usepackage{dirtytalk}
\usepackage{threeparttable}

\shorttitle{Identification of High-Redshift Galaxy Overdensities in GOODS-N and GOODS-S}
\shortauthors{Helton et al.}

\graphicspath{{./}{figures/}}

\begin{document}

\title{Identification of High-Redshift Galaxy Overdensities in GOODS-N and GOODS-S}

%% Authors of the paper
\author[0000-0003-4337-6211]{Jakob M. Helton}
\affiliation{Steward Observatory, University of Arizona, 933 N. Cherry Ave., Tucson, AZ 85721, USA}

\author[0000-0002-4622-6617]{Fengwu Sun}
\affiliation{Steward Observatory, University of Arizona, 933 N. Cherry Ave., Tucson, AZ 85721, USA}

\author[0000-0001-5962-7260]{Charity Woodrum}
\affiliation{Steward Observatory, University of Arizona, 933 N. Cherry Ave., Tucson, AZ 85721, USA}

\author[0000-0003-4565-8239]{Kevin N. Hainline}
\affiliation{Steward Observatory, University of Arizona, 933 N. Cherry Ave., Tucson, AZ 85721, USA}

\author[0000-0001-9262-9997]{Christopher N. A. Willmer}
\affiliation{Steward Observatory, University of Arizona, 933 N. Cherry Ave., Tucson, AZ 85721, USA}

\author[0000-0002-7893-6170]{Marcia J. Rieke}
\affiliation{Steward Observatory, University of Arizona, 933 N. Cherry Ave., Tucson, AZ 85721, USA}

\author[0000-0002-7893-6170]{George H. Rieke}
\affiliation{Steward Observatory, University of Arizona, 933 N. Cherry Ave., Tucson, AZ 85721, USA}

\author[0000-0002-8909-8782]{Stacey Alberts}
\affiliation{Steward Observatory, University of Arizona, 933 N. Cherry Ave., Tucson, AZ 85721, USA}

\author[0000-0002-2929-3121]{Daniel J. Eisenstein}
\affiliation{Center for Astrophysics $|$ Harvard \& Smithsonian, 60 Garden St., Cambridge MA 02138, USA}

\author[0000-0002-8224-4505]{Sandro Tacchella}
\affiliation{Kavli Institute for Cosmology, University of Cambridge, Madingley Road, Cambridge, CB3 OHA, UK}
\affiliation{Cavendish Laboratory, University of Cambridge, 19 JJ Thomson Avenue, Cambridge CB3 0HE, UK}

\author[0000-0002-4271-0364]{Brant Robertson}
\affiliation{Department of Astronomy and Astrophysics, University of California, Santa Cruz, 1156 High Street, Santa Cruz, CA 95064, USA}

\author[0000-0002-9280-7594]{Benjamin D. Johnson}
\affiliation{Center for Astrophysics $|$ Harvard \& Smithsonian, 60 Garden St., Cambridge MA 02138, USA}

\author[0000-0003-0215-1104]{William M. Baker}
\affiliation{Kavli Institute for Cosmology, University of Cambridge, Madingley Road, Cambridge, CB3 OHA, UK}
\affiliation{Cavendish Laboratory, University of Cambridge, 19 JJ Thomson Avenue, Cambridge CB3 0HE, UK}

\author[0000-0003-0883-2226]{Rachana Bhatawdekar}
\affiliation{European Space Agency (ESA), European Space Astronomy Centre (ESAC), Camino Bajo del Castillo s/n, 28692 Villanueva de la Cañada, Madrid, Spain}

\author[0000-0002-8651-9879]{Andrew J. Bunker}
\affiliation{Department of Physics, University of Oxford, Denys Wilkinson Building, Keble Road, Oxford OX1 3RH, UK}

\author[0000-0002-2178-5471]{Zuyi Chen}
\affiliation{Steward Observatory, University of Arizona, 933 N. Cherry Ave., Tucson, AZ 85721, USA}

\author[0000-0003-1344-9475]{Eiichi Egami}
\affiliation{Steward Observatory, University of Arizona, 933 N. Cherry Ave., Tucson, AZ 85721, USA}

\author[0000-0001-7673-2257]{Zhiyuan Ji}
\affiliation{Steward Observatory, University of Arizona, 933 N. Cherry Ave., Tucson, AZ 85721, USA}

\author[0000-0002-4985-3819]{Roberto Maiolino}
\affiliation{Kavli Institute for Cosmology, University of Cambridge, Madingley Road, Cambridge, CB3 OHA, UK}
\affiliation{Cavendish Laboratory, University of Cambridge, 19 JJ Thomson Avenue, Cambridge CB3 0HE, UK}
\affiliation{Department of Physics and Astronomy, University College London, Gower Street, London WC1E 6BT, UK}

\author[0000-0002-4201-7367]{Chris Willott}
\affiliation{NRC Herzberg, 5071 West Saanich Rd, Victoria, BC V9E 2E7, Canada}

\author[0000-0002-7595-121X]{Joris Witstok}
\affiliation{Kavli Institute for Cosmology, University of Cambridge, Madingley Road, Cambridge, CB3 OHA, UK}
\affiliation{Cavendish Laboratory, University of Cambridge, 19 JJ Thomson Avenue, Cambridge CB3 0HE, UK}

\correspondingauthor{Jakob M. Helton}
\email{jakobhelton@arizona.edu}

%% Abstract of the paper.

\begin{abstract}

We conduct a systematic search for high-redshift galaxy overdensities at $4.9 < z_{\,\mathrm{spec}} < 8.9$ in both the GOODS-N and GOODS-S fields using JWST/NIRCam imaging from JADES and JEMS in addition to JWST/NIRCam wide field slitless spectroscopy from FRESCO. High-redshift galaxy candidates are identified using HST+JWST photometry spanning $\lambda = 0.4-5.0\ \mu\mathrm{m}$. We confirmed the redshifts for roughly a third of these galaxies using JWST spectroscopy over $\lambda = 3.9-5.0\ \mu\mathrm{m}$ through identification of  either $\mathrm{H} \alpha$ or $\left[\mathrm{OIII}\right]\lambda5008$ around the best-fit photometric redshift. The rest-UV magnitudes and continuum slopes of these galaxies were inferred from the photometry: the brightest and reddest objects appear in more dense environments and thus are surrounded by more galaxy neighbors than their fainter and bluer counterparts, suggesting accelerated galaxy evolution within overdense environments. We find $17$ significant ($\delta_{\mathrm{gal}} \geq 3.04$, $N_{\mathrm{galaxies}} \geq 4$) galaxy overdensities across both fields ($7$ in GOODS-N and $10$ in GOODS-S), including the two highest redshift spectroscopically confirmed galaxy overdensities to date at $\left< z_{\mathrm{\,spec}} \right> = 7.954$ and $\left< z_{\mathrm{\,spec}} \right> = 8.222$ (representing densities around $\sim 6$ and $\sim 12$ times that of a random volume). We estimate the total halo mass of these large-scale structures to be $11.5 \leq \mathrm{log}_{10}\left(M_{\mathrm{halo}}/M_{\odot}\right) \leq 13.4$ using an empirical stellar mass to halo mass relation, which are likely underestimates as a result of incompleteness. These protocluster candidates are expected to evolve into massive galaxy clusters with $\mathrm{log}_{10}\left(M_{\mathrm{halo}}/M_{\odot}\right) \gtrsim 14$ by $z = 0$. % Follow-up observations of these extreme environments with JWST and/or ALMA will reveal important details of the galaxy and cluster assembly process in the early Universe. 

\end{abstract}

%% Defines the keywords.
\keywords{Early universe (435); Galaxy evolution (594); Galaxy formation (595); \\ High-redshift galaxies (734); High-redshift galaxy clusters (2007)} 

%% Start of section one.
\section{Introduction}
\label{SectionOne}

The large-scale distribution of galaxies in the local Universe consists of filaments, nodes, voids, and walls: the ``cosmic web'' \citep{Geller:1989}. Galaxy clusters inhabit the nodes of the cosmic web and represent the most extreme matter overdensities allowed by the standard cosmological paradigm of hierarchical structure formation \citep{White:1978}. Recently, observational and theoretical progress has been made in tracing the evolution of galaxy clusters back in time \citep[for reviews, see][]{Kravtsov:2012, Alberts:2022}, up to and across the transition to their uncollapsed predecessors \citep[so-called ``protoclusters''; for a review, see][]{Overzier:2016}. Still, these initial structures in the early Universe, which eventually evolved into the galaxy clusters seen today, are not as well understood.

\citet[][]{Chiang:2017} investigated the internal evolution of the star formation and mass assembly of clusters with a set of $N$-body simulations and semi-analytic models, identifying three distinct epochs during the history of cluster formation. The first of these epochs occurs between $z \gtrsim 10$ and $z \sim 5$, where galaxy growth in protoclusters appears to begin in an ``inside-out'' manner. Although the cores only represent a small amount of the protocluster volume at these redshifts, they initially dominated the star formation rate due to the higher mass accretion rate of these massive halos. The second of these epochs occurs between $z \sim 5$ and $z \sim 1.5$, where there exists an extended and prolonged period of star formation, contributing roughly $65\%$ of the total stellar mass within present-day clusters. Some protoclusters could already contain massive cores near the end of this epoch, which would be the first regions to exhibit widespread galaxy quenching and the build-up of the intracluster medium (ICM). Finally, the third of these epochs occurs between $z \sim 1.5$ and $z \sim 0$, where the fraction of stellar mass within the now collapsed cluster is dictated by the rate of infalling galaxies and progressive quenching. The details of the (proto)cluster assembly process are largely erased as a result of relaxation and virialization during this epoch, requiring us to look to higher redshifts for this information. 

The current challenges in understanding protoclusters are a result of (1) their rarity and (2) their composition (consisting of fewer galaxies within more complex dark matter halos that are yet to be virialized), which make obtaining statistical samples of these structures difficult. Observations and systematic searches for galaxy clusters have typically relied on the fact that these objects are virialized (unlike protoclusters), requiring emission from the ICM via X-rays \citep[e.g.,][]{Rosati:2002, Rosati:2004, Rosati:2009, Gobat:2011, Wang:2016} or at sub-millimeter wavelengths via the Sunyaev-Zeldovich (SZ) effect \citep[][]{Staniszewski:2009, Bleem:2015, Planck:2016, Huang:2020, Hilton:2021}. However, current X-ray and sub-millimeter observations are not sensitive enough to properly identify galaxy clusters or protoclusters beyond $z = 2-3$ \citep[e.g.,][]{Tozzi:2022, DiMascolo:2023}.

Alternatively, observations and systematic searches for galaxy protoclusters have typically involved exploring large optical and infrared sky maps for overdensities of red galaxies \citep[e.g.,][]{Clements:2014, Planck:2016a, Greenslade:2018, Martinache:2018}, Lyman-break galaxies \citep[LBGs; e.g.,][]{Steidel:1998, Steidel:2000, Miley:2004, Lee:2014, Toshikawa:2018}, and/or Lyman-alpha emitters \citep[LAEs; e.g.,][]{Shimasaku:2003, Ouchi:2005, Steidel:2005, Dey:2016, Harikane:2019}. Successful methods usually perform these searches around rare and luminous sources such as quasars \citep[QSOs; e.g.,][]{Wold:2003, Stiavelli:2005, Kashikawa:2007, Stevens:2010, Morselli:2014}, high-redshift radio galaxies \citep[HzRGs; e.g.,][]{Pascarelle:1996, LeFevre:1996, Pentericci:2000, Venemans:2004, Venemans:2005, Venemans:2007, Wylezalek:2013}, and sub-millimeter galaxies \citep[SMGs; e.g.,][]{Chapman:2009, Riechers:2014, Umehata:2015, Oteo:2018, Long:2020}. Yet, it has remained difficult to obtain large, representative samples of galaxies at the redshifts necessary to study the earliest phases of the cluster assembly process. 

With the launch of the James Webb Space Telescope (JWST), it is now possible to obtain large samples of rest-optical spectra for high-redshift star-forming galaxies with remarkable efficiency and completeness using data from the Near-Infrared Camera \citep[NIRCam;][]{Rieke:2005, Rieke:2023a}. The powerful combination of deep imaging and wide field slitless spectroscopy (WFSS) provided by JWST/NIRCam have already revealed large-scale galaxy overdensities around the ultraluminous QSOs J0100+2802 at $z = 6.3$ \citep{Kashino:2023} and J0305–3150 at $z = 6.6$ \citep{Wang:2023}, the well-known dusty star-forming galaxy (DSFG) HDF850.1 at $z = 5.2$ \citep{Herard-Demanche:2023, Sun:2023b}, and in a blank region of the sky near the Hubble Ultra Deep Field (HUDF) at $z = 5.4$ \citep{Helton:2023}. Furthermore, at even higher redshifts, multi-object spectroscopy from the Near-Infrared Spectrograph \citep[NIRSpec;][]{Jakobsen:2022, Ferruit:2022} have already revealed an extreme galaxy overdensity around the most distant DSFG SPT$0311-58$ at $z = 6.9$ \citep{Arribas:2023}, the highest redshift spectroscopically confirmed protocluster to date at $z = 7.9$ \citep[][]{Morishita:2023}, and the candidate protocluster core surrounding GN-z11 at $z = 10.6$ \citep[][]{Scholtz:2023}. These results demonstrate the transformative capability of JWST in studying dark matter halo assembly and galaxy formation at very early cosmic times: it is clear that we have entered a new era of studying the large-scale environment of the early Universe.

Here we conduct a systematic search for galaxy overdensities (or protocluster candidates) in both the GOODS-N and GOODS-S fields at $4.9 < z_{\,\mathrm{spec}} < 8.9$ using JWST/NIRCam imaging and WFSS. This paper proceeds as follows. In Section~\ref{SectionTwo}, we describe the various data and observations that are used in our analysis. In Section~\ref{SectionThree}, we present our analysis and results. In Section~\ref{SectionFour}, we summarize our findings and their implications for galaxy evolution in the early Universe. We adopt the protocluster definition from \citet[][]{Alberts:2022}, as extended overdensities at $z \geq 2$ which will collapse into a galaxy cluster by $z = 0$. All magnitudes are in the AB system \citep{Oke:1983}. Uncertainties are quoted as $68\%$ confidence intervals, unless otherwise stated. Throughout this work, we report wavelengths in vacuum and adopt the standard flat $\Lambda$CDM cosmology from Planck18 with $H_{0} = 67.4\ \mathrm{km/s/Mpc}$ and $\Omega_{m} = 0.315$ \citep[][]{Planck:2020}.

%% Start of section two.
\section{Observations}
\label{SectionTwo}

In this work, we use deep optical imaging from the Advanced Camera for Surveys (ACS) on the Hubble Space Telescope (HST) alongside deep infrared imaging and WFSS from JWST/NIRCam in both of the Great Observatories Origins Deep Survey \citep[GOODS;][]{Giavalisco:2004} fields. The photometric data from HST/ACS and JWST/NIRCam are described in Section~\ref{SectionTwoOne} while the spectroscopic data from JWST/NIRCam are described in Section~\ref{SectionTwoTwo}.

\subsection{Photometric Data}
\label{SectionTwoOne}

Our photometric data consist of: (1) deep optical imaging taken with HST/ACS in five photometric bands (F435W, F606W, F775W, F814W, and F850LP) and (2) deep infrared imaging taken with JWST/NIRCam in fifteen photometric bands (F070W, F090W, F115W, F150W, F182M, F200W, F210M, F277W, F335M, F356W, F410M, F430M, F444W, F460M, and F480M).

The HST/ACS mosaics used here were produced as part of the Hubble Legacy Fields (HLF) project v2.0 \citep[][]{Illingworth:2016, Whitaker:2019}. The JWST/NIRCam data were obtained by the JWST Advanced Deep Extragalactic Survey \citep[JADES;][]{Eisenstein:2023} and the JWST Extragalactic Medium-band Survey \citep[JEMS;][]{Williams:2023}. The JADES observations consist of a deep mosaic with nine filters (F090W, F115W, F150W, F200W, F277W, F335M, F356W, F410M, and F444W) and a medium mosaic with eight filters (F090W, F115W, F150W, F200W, F277W, F356W, F410M, and F444W) in both of the GOODS fields. The total survey area of JADES in GOODS-N is roughly $57$ square arcminutes whereas the total survey area in GOODS-S is roughly $77$ square arcminutes. The JEMS observations consist of two $2.2^{\prime} \times 2.2^{\prime}$ regions in GOODS-S with five filters (F182M, F210M, F430M, F460M, and F480M), which all lie in JADES coverage. For all of the subsequent analyses, we do not require any of our objects to have JEMS observations, but we use these data when available. For some regions of the sky, there are nineteen filters that are covered in total, but the majority of regions are only covered by thirteen or fourteen filters. 

A detailed description of the JWST/NIRCam imaging data reduction and mosaicing is outlined in \citet[][]{Tacchella:2023b} and will be presented in detail in a forthcoming paper from the JADES Collaboration (Tacchella et al., in preparation). We briefly summarize here the main steps of the reduction and mosaicing process. The data are initially processed with the standard \texttt{JWST Calibration Pipeline}\footnote{\href{https://github.com/spacetelescope/jwst}{https://github.com/spacetelescope/jwst}\label{JWST_Pipeline}} \citep[v1.9.6;][]{JWST_Pipeline_v1p9p6}. Customized steps are included to aid in the removal of ``1/f'' noise, ``wisp'' artifacts, ``snowball'' artifacts, and persistence from previous observations \citep[see also][]{Rigby:2023}. The JWST Calibration Reference Data System (CRDS; v11.16.21) context map is used (jwst\_1084.pmap), including the flux calibration for JWST/NIRCam from Cycle 1. The sky background is modeled and removed using the BackGround2D class from \texttt{photutils} \citep[][]{Bradley:2022}. Finally, the image mosaics for each of the JWST/NIRCam filters are registered to the \textsc{Gaia} DR3 frame \citep[][]{GaiaDR3} and resampled onto the same world coordinate system (WCS) with a $30\,\mathrm{mas/pixel}$ grid. Assuming circular apertures with diameters of $0.2^{\prime\prime}$, the $5\sigma$ point-source detection limit in the F277W filter is $m \approx 31\,\mathrm{AB\,mag}$ and $m \approx 29\,\mathrm{AB\,mag}$ for the deepest and shallowest regions of JADES, respectively. 

A detailed description of the JWST/NIRCam source detection is outlined in \citet[][]{Robertson:2023} and will be presented in detail in another forthcoming paper from the JADES Collaboration (Robertson et al., in preparation). We briefly summarize here the main steps of the source detection process. Six image mosaics (F200W, F277W, F335M, F356W, F410M, and F444W) are initially stacked using the corresponding error images and inverse-variance weighting to produce a single detection image. Within this detection image, we construct a source catalog by selecting contiguous regions of greater than five pixels with signal-to-noise ratios $\mathrm{S/N} > 3$ and applying a standard Source Extractor \citep[\texttt{SExtractor};][]{SourceExtractor} deblending algorithm \citep[][]{Bradley:2022}. Finally, we perform forced convolved photometry at the source centroids in all of the HST/ACS and JWST/NIRCam photometric bands, assuming both circular apertures with diameters of $0.2^{\prime\prime}$ as well as elliptical Kron apertures with \texttt{parameter = 1.2} (i.e., Kron small) and \texttt{parameter = 2.5} (i.e., Kron large). To correct for potential missing light, we rescale the Kron small photometry by the flux ratio of Kron large to Kron small in the F444W filter. This methodology was chosen since the smaller elliptical Kron apertures reduce the background noise associated with the use of the larger elliptical Kron apertures. Aperture corrections are applied using model point spread functions (PSFs) from the \texttt{TinyTim} \citep[][]{Krist:2011} package for HST/ACS and the \texttt{WebbPSF} \citep[][]{Perrin:2014} package for JWST/NIRCam, assuming point source morphologies. Uncertainties are estimated by placing random apertures across regions of the image mosaics to compute a flux variance \citep[e.g.,][]{Labbe:2005, Quadri:2007, Whitaker:2011}, which are summed in quadrature with the associated Poisson uncertainty for each detected source. To account for differing depths across the fields, these random apertures are collected into percentiles based on the exposure time at each location, which are used to determine the sky-noise contribution to the flux uncertainty for each individual object.

\subsection{Spectroscopic Data}
\label{SectionTwoTwo}

Our spectroscopic data consist of WFSS taken with JWST/NIRCam in the F444W filter ($\lambda = 3.9-5.0\ \mu\mathrm{m}$). These data were obtained by the First Reionization Epoch Spectroscopic COmplete Survey \citep[FRESCO;][]{Oesch:2023}. The FRESCO observations cover a $7.4^{\prime} \times 8.4^{\prime}$ area using the row-direction grisms on both modules of JWST/NIRCam (Grism R; $R \approx 1600$) in both of the GOODS fields. The total overlapping area between the JADES and FRESCO footprints is roughly $35$ square arcminutes in GOODS-N and roughly $46$ square arcminutes in GOODS-S.

A detailed description of the JWST/NIRCam grism data reduction will be described in a forthcoming paper from the JADES Collaboration (Sun et al., in preparation), which is similar to those presented in \citet[][]{Sun:2023a} and \citet[][]{Helton:2023}. We briefly summarize here the main steps of the reduction process\footnote{\href{https://github.com/fengwusun/nircam_grism}{https://github.com/fengwusun/nircam\_grism}\label{FengwuSun_Grism}}. The data are initially processed with the standard \texttt{JWST Calibration Pipeline}\textsuperscript{\ref{JWST_Pipeline}} \citep[v1.11.2;][]{JWST_Pipeline_v1p11p2}. We place the rate files into the WCS, perform flat-fielding, and subtract out the sigma-clipped median sky background from each individual exposure after the ``ramp-to-slope'' fitting in the calibration pipeline. A median-filtering technique is utilized to subtract out any remaining continuum or background on a row-by-row basis following the methodology of \citet[][]{Kashino:2023}. This produces emission line maps for each grism exposure that are devoid of continuum. Although this median-filtering technique is able to properly remove the continuum contamination, it occasionally over-subtracts signal in the spectral regions immediately surrounding the brightest emission lines. \textcolor{black}{This issue will be further discussed and mitigated in a forthcoming paper from the JADES Collaboration (Sun et al., in preparation).} We further remove the ``1/f'' noise using the \texttt{tshirt/roeba} algorithm\footnote{\href{https://github.com/eas342/tshirt}{https://github.com/eas342/tshirt}} in both the row and column directions. SW parallel observations were conducted in two photometric bands (F182M and F210M) and are used for both astrometric and wavelength calibration of the long wavelength (LW) spectroscopic data. We use the spectral tracing and grism dispersion models that were produced using the JWST/NIRCam commissioning data of the Large Magellanic Cloud \citep[LMC;][]{Sun:2023a}. We additionally make use of the flux calibration models that were produced from the JWST/NIRCam Cycle-1 absolute flux calibration observations \cite[][]{Gordon:2022}.

%% Start of section three.
\section{Analysis \& Results}
\label{SectionThree}

Using the observations from Section~\ref{SectionTwo}, we perform various analyses and present the results here. Photometric redshifts and the selection of our photometric sample are described in Section~\ref{SectionThreeOne}. Spectroscopic redshifts and the selection of the spectroscopic sample are described in Section~\ref{SectionThreeTwo}. Galaxy overdensity values and the identification of high-redshift galaxy overdensities are described in Section~\ref{SectionThreeThree}, where each of these structures is discussed in detail. Detailed physical modeling of the rest-ultraviolet (UV) photometry and the comparison of these properties with their galaxy overdensity values are described in Section~\ref{SectionThreeFour}. Detail physical modeling of the stellar populations and the presentation of the star-forming main sequence are described in Section~\ref{SectionThreeFive}. Estimating the total dark matter halo mass for each of the high-redshift galaxy overdensities and placing these structures in the context of theoretical predictions are described in Section~\ref{SectionThreeSix}. 

\subsection{Photometric Redshifts}
\label{SectionThreeOne}

Using the rescaled Kron small photometry described in Section~\ref{SectionTwoOne}, we measure photometric redshifts with the template-fitting code \texttt{EAZY} \citep[][]{Brammer:2008}. \texttt{EAZY} makes use of a chi-square ($\chi^{2}$) minimization technique to model the broadband spectral energy distributions (SEDs) of galaxies using linear combinations of galaxy templates. We utilize sixteen templates in total to perform the fitting, which includes the nine \texttt{EAZY} ``v1.3'' templates, two additional templates for simple stellar populations with ages of $5\,\mathrm{Myr}$ and $25\,\mathrm{Myr}$, and five more templates with strong nebular continuum emission. A detailed description of the photometric redshift estimation is provided in \citet[][]{Helton:2023} and \citet[][]{Hainline:2023}. We briefly describe here the main steps of the photometric redshift process.

Photometric offsets are iteratively calculated from the \texttt{EAZY} templates compared to the true JWST/NIRCam photometry to minimize the effects of imperfect photometric calibration, using a sample of objects with $5 < \mathrm{S/N} < 20$ in F200W. These photometric offsets are relatively small and are subsequently applied to the entire photometric catalog. We choose not to adopt any apparent magnitude priors, but we do make use of the template error file \textsc{``template\_error.v2.0.zfourge''}. The primary measurements used here are the \texttt{EAZY} ``$z_{\mathrm{a}}$'' and ``$z_{\mathrm{peak}}$'' redshifts. The former corresponds to the fit where the likelihood is maximized ($\chi^2$ is minimized), while the latter corresponds to the weighted average of the fits where the probability is maximized (the probability is equal to the integral of the likelihood). We allow \texttt{EAZY} to fit across the redshift range of $z = 0.2 - 22$ with a redshift step size of $\Delta z = 0.01$.

To perform an accurate and efficient targeted emission line search, we require a sample of relatively bright objects (since these are the only objects for which we expect to detect an emission line) with narrow photometric redshift constraints (since this allows for spectroscopic confirmation using only a single line detection). Following the methodology of \citet[][]{Helton:2023}, we select galaxies that satisfy the following selection criteria:
\begin{enumerate}
    \item Both of the best-fit \texttt{EAZY} photometric redshifts must be between four and a half and nine and a half (i.e., $4.5 < z_{\mathrm{a}} < 9.5$ and $4.5 < z_{\mathrm{peak}} < 9.5$).
    \item The apparent magnitude in F444W assuming elliptical Kron apertures with \texttt{parameter = 2.5} must be at least $0.5\ \mathrm{AB\,mag}$ brighter than the shallowest region of JADES (i.e., $m < 28.5\ \mathrm{AB\,mag}$).
    \item The first \texttt{EAZY} confidence interval must be \textcolor{black}{less than twenty percent of one plus the best-fit \texttt{EAZY} photometric redshifts (i.e., $\Delta z_{\,1} < 0.2\left[1 + z_{\mathrm{a}}\right]$ and $\Delta z_{\,1} < 0.2\left[1 + z_{\mathrm{peak}}\right]$)}. This confidence interval is defined to be the difference between the 16th and 84th percentiles of the photometric redshift posterior distribution, which is roughly equal to two times the standard deviation ($\approx \pm 1\sigma$). 
    \item The second \texttt{EAZY} confidence interval must be \textcolor{black}{less than forty percent of one plus the best-fit \texttt{EAZY} photometric redshifts (i.e., $\Delta z_{\,2} < 0.4\left[1 + z_{\mathrm{a}}\right]$ and $\Delta z_{\,2} < 0.4\left[1 + z_{\mathrm{peak}}\right]$)}. This confidence interval is defined to be the difference between the 5th and 95th percentiles of the photometric redshift posterior distribution, which is roughly equal to four times the standard deviation ($\approx \pm 2\sigma$). 
\end{enumerate}

\begin{figure}
    \centering
    \includegraphics[width=1.0\linewidth]{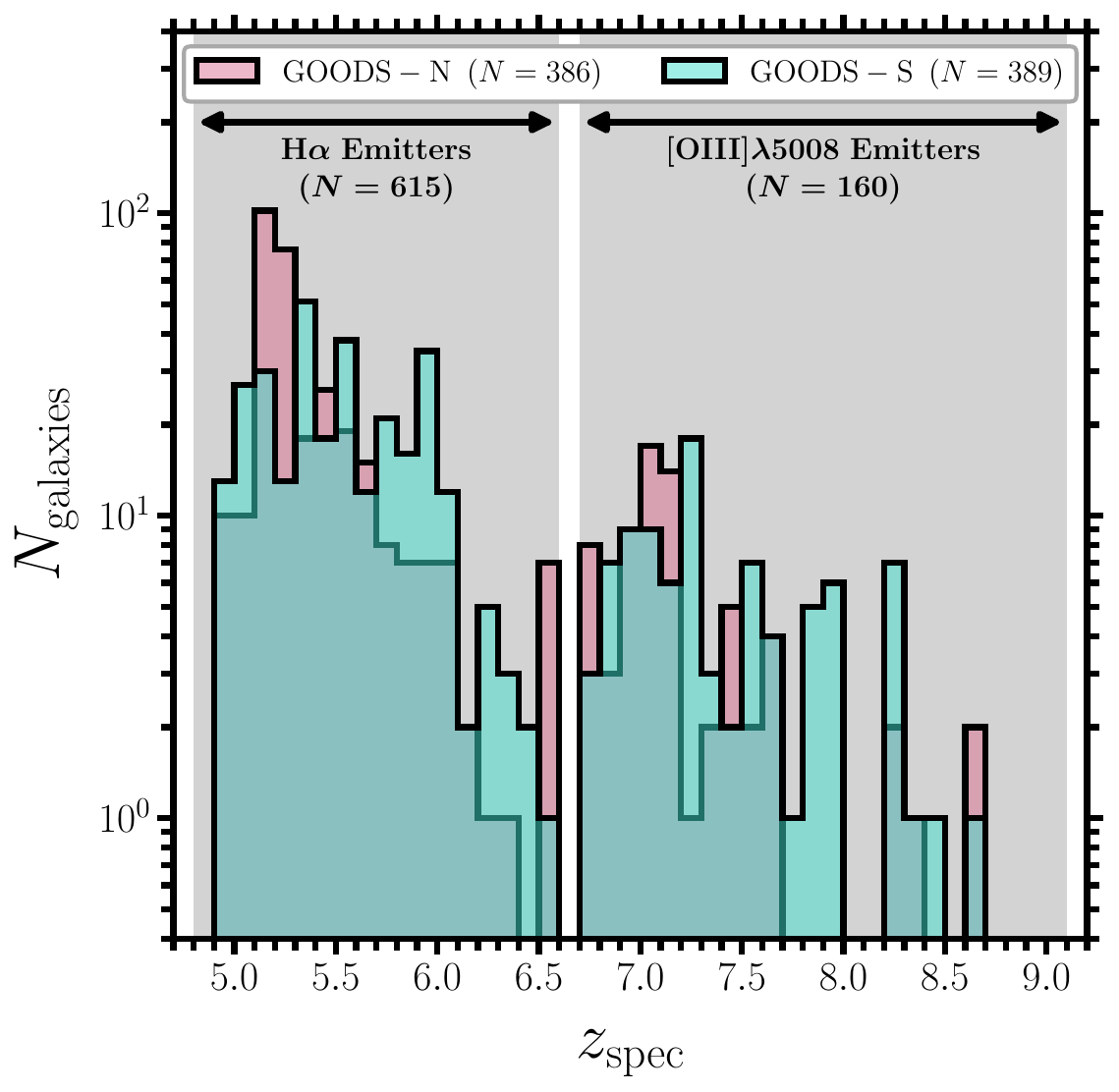}
    \caption{Histograms showing the distribution of spectroscopic redshifts for the final spectroscopic sample described in Section~\ref{SectionThreeTwo}. The pink shaded regions represent the galaxies from GOODS-N while the blue shaded regions represent the galaxies from GOODS-S. This sample includes objects across the redshift range $4.9 < z_{\,\mathrm{spec}} < 8.9$ with $\mathrm{H}\alpha$ or $\mathrm{[OIII]}\lambda5008$ detections. \label{fig:Figure_01}}
\end{figure}

These cuts produced an initial photometric sample of relatively bright objects with narrow photometric redshift constraints, which means we are likely missing some additional subset of objects with relatively unconstrained photometric redshifts (e.g., DSFGs and/or massive quiescent galaxies). We subsequently visually inspected the entirety of this photometric sample to remove obvious nonphysical data artifacts, including extended diffraction spikes from bright stars and hot pixels caused by cosmic rays. Low-redshift, extended, dusty sources masquerading as high-redshift dropouts are additionally removed. Following this visual inspection, we are left with a final photometric sample of \textcolor{black}{$N = 2656$} galaxies in total across the redshift range $4.5 < z_{\mathrm{phot}} < 9.5$ ($N = 1228$ objects in GOODS-N, \textcolor{black}{$N = 1428$} objects in GOODS-S). This range in photometric redshifts was chosen to encompass all possible $\mathrm{H}\alpha$ or $\mathrm{[OIII]}\lambda5008$ emitters.

\subsection{Spectroscopic Redshifts}
\label{SectionThreeTwo}

Using the continuum-subtracted grism images from Section~\ref{SectionTwoTwo}, we extract two-dimensional (2d) grism spectra for all of the galaxies that are part of the final photometric sample described in Section~\ref{SectionThreeOne}. We further optimally extract \citep{Horne:1986} one-dimensional (1d) grism spectra using the observed surface brightness profile in the F444W imaging. Peaks are subsequently identified using an automatic line fitting routine, which assumes various bin sizes (integer units of $\mathrm{nm}$ from one to eight). After removing duplicated peaks identified with different bin sizes, each peak that is detected with $\mathrm{S/N} > 3$ is then fit with a Gaussian line profile. For each line that is detected with $\mathrm{S/N} > 3$, we tentatively assign a line identification of either $\mathrm{H}\alpha$ or $\mathrm{[OIII]}\lambda5008$: whichever one minimizes the difference between the best-fit photometric redshift and the tentative spectroscopic redshift. For example, if a line were detected at $\lambda = 4.2\,\mu\mathrm{m}$ and the best-fit photometric redshift is $z_{\mathrm{phot}} = 5.8$, then the initial line identification would be $\mathrm{H}\alpha$, since the predicted wavelength of this line would be at $\lambda = 4.5\,\mu\mathrm{m}$, which is closer to the observed wavelength than the predicted wavelength of $\mathrm{[OIII]}\lambda5008$ ($\lambda = 3.4\,\mu\mathrm{m}$). 

Visual inspection is performed on each of these tentative spectroscopic redshift solutions to remove spurious detections caused by either noise or contamination from nearby sources along the dispersion direction. Galaxies that pass our visual inspection are determined to have secure line detections. We optimally re-extract the 1d spectra \citep[][]{Horne:1986} for the sources with secure line detections using the observed spectral line profile along the spatial direction (perpendicular to the dispersion direction). We once again fit the detected peaks with Gaussian line profiles. The typical absolute uncertainties of our spectroscopic redshifts are $\Delta z_{\,\mathrm{spec}} = 0.001$, which are derived from the grism wavelength calibration uncertainties. Following this procedure, we are left with a final spectroscopic sample of $N = 775$ galaxies in total across the redshift range $4.9 < z_{\,\mathrm{spec}} < 8.9$ ($N = 386$ of these objects are in GOODS-N while $N = 389$ of these objects are in GOODS-S). Within this sample, there are $N = 615$ galaxies with $\mathrm{H}\alpha$ detections across the redshift range $4.9 < z_{\,\mathrm{spec}} < 6.6$ in addition to $N = 160$ galaxies with $\mathrm{[OIII]}\lambda5008$ detections at $6.7 < z_{\,\mathrm{spec}} < 8.9$. 

\begin{figure}
    \centering
    \includegraphics[width=1.0\linewidth]{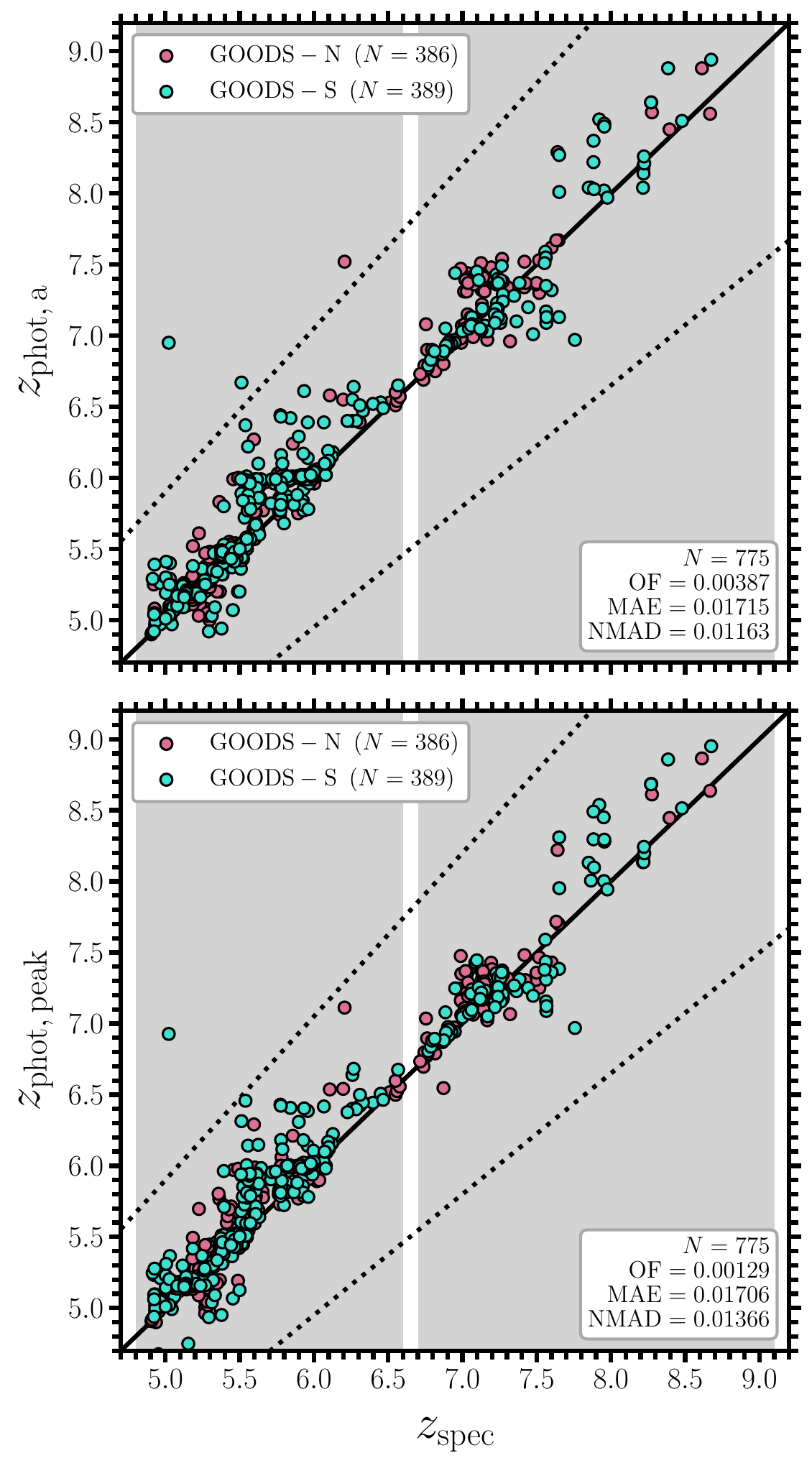}
    \caption{Photometric redshifts versus spectroscopic redshifts for the final spectroscopic sample described in Section~\ref{SectionThreeTwo}. The upper (lower) panel adopts the \texttt{EAZY} ``$z_{\mathrm{a}}$'' (``$z_{\mathrm{peak}}$'') values as the photometric redshifts, which are described in Section~\ref{SectionThreeOne}. The pink (blue) points represent the galaxies from GOODS-N (GOODS-S). The solid black line represents the one-to-one relation while the dotted black lines represent the criteria for being selected as an outlier. Galaxies typically appear above the one-to-one relation since redshifts derived from photometry are systematically over-predicted when compared to those derived from spectroscopy. Some relevant statistical quantities are given in the legends, including: the outlier fraction (OF), the mean absolute error (MAE), and the normalized median absolute deviation (NMAD). These results show that the redshifts derived from the photometry agree remarkably well with those derived from the spectroscopy, although we should note that the initial photometric sample required galaxies to be relatively bright with narrow photometric redshift constraints. \label{fig:Figure_02}}
\end{figure}

Figure~\ref{fig:Figure_01} shows the distribution of spectroscopic redshifts for the final spectroscopic sample, where the pink shaded regions represent galaxies from GOODS-N and the blue shaded regions represent galaxies from GOODS-S. Similarly, Figure~\ref{fig:Figure_02} shows the comparison of photometric and spectroscopic redshifts, where again the pink and blue points represent galaxies from GOODS-N and GOODS-S, respectively. The solid black line represents the one-to-one relation between these two redshifts while the dotted black lines represent the criteria for being selected as an outlier. At these redshifts, galaxies typically appear above the one-to-one relation since photometric redshifts are systematically over-predicted when compared to spectroscopic redshifts. This result has been discussed recently in the literature \citep[e.g.,][]{Arrabal-Haro:2023, Fujimoto:2023, Hainline:2023} and has typically been attributed to differences between the observed high-redshift galaxy SEDs and the templates used to model these galaxies. Further explorations into the possible origin of this offset will be presented in a forthcoming paper from the JADES Collaboration (Sun et al., in preparation), where damped Lyman-$\alpha$ absorption is required to properly model the galaxies with $\mathrm{[OIII]}\lambda5008$ detections  at $6.7 < z_{\,\mathrm{spec}} < 8.9$. The following relevant statistical quantities are given in the legends: the total number of galaxies ($N$), the outlier fraction (OF; the fraction of objects that satisfy Equation~\ref{EquationTwo}), the mean absolute error (MAE; this value is defined in Equation~\ref{EquationThree}), and the median absolute deviation (NMAD; this value is defined in Equation~\ref{EquationFour}).
\begin{equation}
    \label{EquationOne}
    \Delta z = z_{\mathrm{spec}} - z_{\mathrm{phot}}
\end{equation}
\begin{equation}
    \label{EquationTwo}
    \frac{| \Delta z |}{1 + z_{\mathrm{spec}}} > 0.15
\end{equation}
\begin{equation}
    \label{EquationThree}
    \sigma_{\mathrm{MAE}} = \frac{\Sigma_{i = 0}^{N} \lvert \Delta z_{\,i} \rvert}{N}
\end{equation}
\begin{equation}
    \label{EquationFour}
    \sigma_{\mathrm{NMAD}} = 1.48 \times \mathrm{median}\left( \frac{\lvert \Delta z - \mathrm{median}\left[ \Delta z \right] \rvert}{1 + z_{\mathrm{spec}}} \right)
\end{equation}

These statistical quantities indicate the remarkable fidelity of our photometric redshifts. It is important to remember that the initial photometric sample, which was selected in Section~\ref{SectionThreeOne}, was chosen to only include relatively bright galaxies with narrow photometric redshift constraints, so one would expect nearly all of the photometric redshifts to agree with the spectroscopic redshifts. Further validations of the JWST/NIRCam grism redshift and line flux measurements will be presented in a forthcoming paper from the JADES Collaboration (Sun et al., in preparation).

\subsection{Overdensity Identification}
\label{SectionThreeThree}

Following the technique described in \citet[][]{Calvi:2021} and \citet[][]{Helton:2023}, we use a Friends-of-Friends (FoF) algorithm to initially identify high-redshift galaxy overdensities within the final spectroscopic sample described in Section~\ref{SectionThreeOne}. This algorithm iteratively selects structural groupings in three-dimensions (3d; two spatial, one spectral) using only a galaxy's right ascension, declination, and spectroscopic redshift \citep[see also][]{Huchra:1982, Eke:2004, Berlind:2006}. Groups are found by identifying galaxies with projected distances and line-of-sight (LOS) velocity separations that are less than some threshold values ($d_{\mathrm{link}} = 500\,\mathrm{kpc}$ and $\sigma_{\mathrm{link}} = 500\,\mathrm{km/s}$, respectively). These threshold values are referred to as linking parameters and are set as fixed values, for simplicity\footnote{Dynamic linking parameters have been used in the literature \citep[e.g.,][]{Farrens:2011}, although this choice has the consequence of producing a complex selection function, making interpretations more difficult. On the other hand, fixed linking parameters may lead to some redshift dependent selection effects being ignored.}. \textcolor{black}{The groupings identified with this algorithm produce similar results when varying either the projected distance ($d_{\mathrm{link}}$) or the LOS velocity separation ($\sigma_{\mathrm{link}}$) by a factor of two, although future work should look at the applicability of these linking parameters in identifying robust structural groupings at high redshift.}

\begin{table*}
	\caption{A summary of physical quantities for the final sample of galaxy overdensities, described in Section~\ref{SectionThreeThree}.}
	\label{tab:overdensities}
	\begin{threeparttable}
	\makebox[\textwidth]{
	\hspace*{-20mm}
        \begin{tabular}{ccccccc}
		\hline
		\hline
		Name & $\mathrm{R.A.\ (J2000)}$\tnote{a} & $\mathrm{Decl.\ (J2000)}$\tnote{b} & $N_{\mathrm{galaxies}}$\tnote{c} & $\left< z_{\mathrm{\,spec}} \right>$\tnote{d} & $\left< \delta_{\mathrm{gal}} \right>$\tnote{e} & $\mathrm{log}_{10} \left( M_{\mathrm{halo}}/M_{\odot} \right)$\tnote{f} \\
		\hline
            JADES$-$GN$-$OD$-5.191$ & $189.21047$ & $62.22105$ & $103$ & \textcolor{black}{$5.191 \pm 0.002$} & \textcolor{black}{$8.924 \pm 0.378$} & $13.436 \pm 0.039$ \\
            JADES$-$GN$-$OD$-5.194$ & $189.14389$ & $62.30190$ & $8$ & \textcolor{black}{$5.194 \pm 0.001$} & \textcolor{black}{$3.597 \pm 0.540$} & $12.252 \pm 0.163$ \\
            JADES$-$GN$-$OD$-5.269$ & $189.12783$ & $62.24182$ & $14$ & \textcolor{black}{$5.269 \pm 0.001$} & \textcolor{black}{$3.472 \pm 0.155$} & $12.375 \pm 0.074$ \\
            JADES$-$GS$-$OD$-5.386$ & $53.08846$ & $-27.82449$ & $39$ & \textcolor{black}{$5.386 \pm 0.001$} & \textcolor{black}{$9.134 \pm 0.172$} & $13.038 \pm 0.044$ \\
            JADES$-$GS$-$OD$-5.928$ & $53.13527$ & $-27.78893$ & $14$ & \textcolor{black}{$5.928 \pm 0.002$} & \textcolor{black}{$4.923 \pm 0.223$} & $12.503 \pm 0.057$ \\
            JADES$-$GS$-$OD$-6.876$ & $53.14291$ & $-27.78575$ & $4$ & \textcolor{black}{$6.876 \pm 0.003$} & \textcolor{black}{$3.907 \pm 0.506$} & $11.462 \pm 0.122$ \\
            JADES$-$GS$-$OD$-6.906$ & $53.10573$ & $-27.84951$ & $4$ & \textcolor{black}{$6.906 \pm 0.001$} & \textcolor{black}{$5.854 \pm 0.183$} & $11.810 \pm 0.154$ \\
            JADES$-$GN$-$OD$-6.991$ & $189.17527$ & $62.25518$ & $6$ & \textcolor{black}{$6.991 \pm 0.001$} & \textcolor{black}{$6.142 \pm 0.253$} & $11.660 \pm 0.111$ \\
            JADES$-$GN$-$OD$-7.025$ & $189.16789$ & $62.29100$ & $6$ & \textcolor{black}{$7.025 \pm 0.004$} & \textcolor{black}{$4.666 \pm 0.550$} & $11.614 \pm 0.096$ \\
            JADES$-$GS$-$OD$-7.061$ & $53.07610$ & $-27.83756$ & $5$ & \textcolor{black}{$7.061 \pm 0.003$} & \textcolor{black}{$3.866 \pm 0.421$} & $11.632 \pm 0.133$ \\
            JADES$-$GN$-$OD$-7.133$ & $189.12636$ & $62.23430$ & $5$ & \textcolor{black}{$7.133 \pm 0.003$} & \textcolor{black}{$5.247 \pm 1.174$} & $11.852 \pm 0.146$ \\
            JADES$-$GN$-$OD$-7.144$ & $189.25454$ & $62.23888$ & $4$ & \textcolor{black}{$7.144 \pm 0.002$} & \textcolor{black}{$5.086 \pm 0.114$} & $11.834 \pm 0.195$ \\
            JADES$-$GS$-$OD$-7.231$ & $53.17006$ & $-27.78024$ & $5$ & \textcolor{black}{$7.231 \pm 0.004$} & \textcolor{black}{$6.805 \pm 0.848$} & $11.733 \pm 0.136$ \\
            JADES$-$GS$-$OD$-7.265$ & $53.18444$ & $-27.78779$ & $7$ & \textcolor{black}{$7.265 \pm 0.003$} & \textcolor{black}{$6.631 \pm 0.500$} & $11.683 \pm 0.106$ \\
            JADES$-$GS$-$OD$-7.561$ & $53.17739$ & $-27.75186$ & $7$ & \textcolor{black}{$7.561 \pm 0.002$} & \textcolor{black}{$6.465 \pm 0.126$} & $11.909 \pm 0.116$ \\
            JADES$-$GS$-$OD$-7.954$ & $53.09377$ & $-27.86057$ & $4$ & \textcolor{black}{$7.954 \pm 0.001$} & \textcolor{black}{$5.495 \pm 0.106$} & $11.606 \pm 0.148$ \\
            JADES$-$GS$-$OD$-8.220$ & $53.13935$ & $-27.83658$ & $6$ & \textcolor{black}{$8.220 \pm 0.001$} & \textcolor{black}{$10.735 \pm 0.212$} & $11.791 \pm 0.136$ \\
            \hline
	\end{tabular}
        }
	\begin{tablenotes}
	    \footnotesize
            \item \textbf{Notes.}
            \item[a] Mean of the constituent right ascensions.
            \item[b] Mean of the constituent declinations.
            \item[c] Total number of constituent galaxies.
            \item[d] Mean and standard error of the constituent spectroscopic redshifts.
            \item[e] Mean and standard error of the constituent galaxy overdensity values.
            \item[f] Mean and standard deviation of the inferred halo mass as described in Section~\ref{SectionThreeSix}.
        \end{tablenotes}
	\end{threeparttable}
\end{table*}

Following the methodology of \citet{Chiang:2013}, we calculate 3d galaxy overdensities ($\delta_{\mathrm{gal}} = n_{\mathrm{gal}} / \langle n_{\mathrm{gal}} \rangle - 1$), where $n_{\mathrm{gal}}$ is the number density of galaxies and $\langle n_{\mathrm{gal}} \rangle$ is the ensemble average number density of galaxies, which are both estimated individually for each galaxy within the final spectroscopic sample. To calculate these values, we assume a tophat-weighted spherical window. For the former (i.e., $n_{\mathrm{gal}}$), the number density is found by placing an aperture at the location of each individual galaxy. For the latter (i.e., $\langle n_{\mathrm{gal}} \rangle$), the ensemble average number density is found by placing $1000$ random apertures within the JADES and FRESCO footprints. All of the apertures considered here have comoving volumes that are equal to the volumes associated with the $\left( 15\,\mathrm{cMpc} \right)^{3}$ tophat-weighted cubic windows assumed by \citet{Chiang:2013}. We further require these apertures to be within line-of-sight velocities $\lvert \Delta v \rvert \leq 10^{4}\,\mathrm{km/s}$ of each individual galaxy to avoid any potential redshift evolution. This requirement is especially important for redshifts where the JWST/NIRCam Grism R line flux sensitivity is largely dependent on the observed wavelength. We find that the number densities calculated in this way produce similar results when varying the line-of-sight velocity selection by a factor of a few. The uncertainty on the ensemble average is found by taking the standard deviation of the values found from the 1000 random apertures, while the uncertainty on the 3d galaxy overdensity value is found by propagating the uncertainty on the ensemble average with the Poisson uncertainty associated with the galaxy number density for each individual galaxy.

Robust high-redshift galaxy overdensities are initially selected from the groupings identified by the FoF algorithm by requiring a minimum number of constituent galaxies ($N_{\mathrm{galaxies}} \geq 4$) and an average galaxy overdensity value that is larger than the value taken from \citet[][$\delta_{\mathrm{gal}} \geq 3.04$]{Chiang:2013}. This galaxy overdensity value is taken for the $z = 5$ SFR-limited sample ($\mathrm{SFR} > 1\,M_{\odot}/\mathrm{yr}$) assuming tophat-weighted cubic window sizes of $\left( 15\,\mathrm{cMpc} \right)^{3}$, which is the sample that is most similar to our own in terms of selection, and identifies structures within cosmological simulations as protocluster candidates with $80\%$ confidence. For comparison, the SFR detection limit for the galaxies with $\mathrm{H}\alpha$ detections is $\sim 2\,M_{\odot}/\mathrm{yr}$ at $z \approx 5$, assuming the conversion factor from \citet{Kennicutt:2012}. Following this selection, we are left with a final sample of $N = 17$ galaxy overdensities (or protocluster candidates) in total ($N = 7$ of these candidates are in GOODS-N, $N = 10$ of these candidates are in GOODS-S). A summary of observed and inferred properties for these galaxy overdensities are given in Table~\ref{tab:overdensities}. We will now discuss each overdensity individually and highlight any previous discussion of these overdensities in the literature. The reported angular and physical diameters are measured as the maximum separation between the constituent galaxies for a given overdensity.

\input{figure_set}

JADES$-$GN$-$OD$-5.191$ is an overdensity in GOODS-N with $103$ galaxies, making it the largest overdensity in terms of total number of galaxies in either field searched (as well as the largest in terms of inferred halo mass in either field). The average overdensity value is $\delta_{\mathrm{gal}} = 8.924$ across an inner 68th percentile range of $3.588$ and $12.966$. The angular (physical) diameter is $6.6$ arcmin ($15.3$ cMpc). This overdensity was discussed previously in \citet{Walter:2012}, where $13$ spectroscopic sources were found to be associated with an overdensity containing the well-known sub-millimeter galaxy HDF850.1; then in \citet{Arrabal-Haro:2018}, where $55$ photometric sources were found to be related to the same overdensity, although spectroscopic follow-up was needed to confirm the association of these candidates; then in \citet{Calvi:2021}, where an additional $10$ newly confirmed members were found; and finally in \citet{Herard-Demanche:2023} and \citet{Sun:2023b}, where $100$ and $109$ sources were found to be in the vicinity of HDF850.1, respectively (note that these two works adopted different survey areas and definitions for galaxy associations). In addition to JADES$-$GN$-$OD$-5.194$ and JADES$-$GN$-$OD$-5.269$, these overdensities reside in a complex environment with filamentary structures, as previously discussed in both \citet{Herard-Demanche:2023} and \citet{Sun:2023b}. JADES$-$GN$-$OD$-5.194$ contains $8$ galaxies. The average overdensity value is $\delta_{\mathrm{gal}} = 3.597$ across a similar range of $1.644$ and $4.801$. The angular (physical) diameter is $1.7$ arcmin ($3.9$ cMpc). JADES$-$GN$-$OD$-5.269$ contains $14$ galaxies. The average overdensity value is $\delta_{\mathrm{gal}} = 3.472$ across a similar range of $3.141$ and $3.767$. The angular (physical) diameter is $2.4$ arcmin ($5.7$ cMpc).

JADES$-$GS$-$OD$-5.386$ is an overdensity in GOODS-S with $39$ galaxies, making it the second largest overdensity in terms of total number of galaxies in either field searched (as well as the second largest in terms of inferred halo mass in either field). The average overdensity value is $\delta_{\mathrm{gal}} = 9.134$ across an inner 68th percentile range of $8.613$ and $9.867$, making it the most significant overdensity consisting of $\mathrm{H}\alpha$ emitters (at $4.9 < z_{\,\mathrm{spec}} < 6.6$) in terms of average overdensity value between either field. The angular (physical) diameter is $3.6$ arcmin ($8.6$ cMpc). This overdensity was discussed previously in \citet{Helton:2023}, where the same $N = 39$ spectroscopic sources were found to be associated with an extreme galaxy overdensity.

JADES$-$GS$-$OD$-5.928$ is an overdensity in GOODS-S with $14$ galaxies. The average overdensity value is $\delta_{\mathrm{gal}} = 4.923$ across an inner 68th percentile range of $3.885$ and $5.506$. The angular (physical) diameter is $3.9$ arcmin ($9.6$ cMpc). \textcolor{black}{A potential large-scale structure was identified in the HUDF at $z \approx 5.9$ by \citet[][]{Stanway:2007} after finding two LAEs at $z = 5.93$, but neither are part of this overdensity.} More recently, one of the sources which make up this overdensity was identified as a LAE at $z = 5.93$ in \citet[][]{Witstok:2023}, where roughly a third of the LAEs that were presented were found to coincide with large-scale galaxy overdensities at $z \approx 5.9$, potentially corresponding to ionized bubbles with sizes of $R_{\mathrm{ion}} = 3.8-6.4$ cMpc.

JADES$-$GS$-$OD$-6.876$ is an overdensity in GOODS-S with $4$ galaxies. The average overdensity value is $\delta_{\mathrm{gal}} = 3.907$ across an inner 68th percentile range of $2.898$ and $4.919$. The angular (physical) diameter is $2.2$ arcmin ($5.7$ cMpc). Similarly, JADES$-$GS$-$OD$-6.906$ contains $4$ galaxies. The average overdensity value is $\delta_{\mathrm{gal}} = 5.854$ across a similar range of $5.595$ and $6.124$. The angular (physical) diameter is $0.4$ arcmin ($0.9$ cMpc). These overdensities have not been discussed previously in the literature.

JADES$-$GN$-$OD$-6.991$ is an overdensity in GOODS-N with $6$ galaxies. The average overdensity value is $\delta_{\mathrm{gal}} = 6.142$ across an inner 68th percentile range of $5.583$ and $6.753$. The angular (physical) diameter is $1.3$ arcmin ($3.3$ cMpc). JADES$-$GN$-$OD$-7.025$ contains $6$ galaxies. The average overdensity value is $\delta_{\mathrm{gal}} = 4.666$ across a similar range of $3.490$ and $5.549$. The angular (physical) diameter is $1.8$ arcmin ($4.6$ cMpc). JADES$-$GN$-$OD$-7.133$ contains $5$ galaxies. The average overdensity value is $\delta_{\mathrm{gal}} = 3.866$ across a similar range of $3.186$ and $7.961$. The angular (physical) diameter is $1.2$ arcmin ($3.1$ cMpc). JADES$-$GN$-$OD$-7.144$ contains $4$ galaxies. The average overdensity value is $\delta_{\mathrm{gal}} = 5.086$ across a similar range of $4.921$ and $5.242$. The angular (physical) diameter is $0.4$ arcmin ($1.1$ cMpc). These overdensities have not been discussed previously in the literature, but they reside in a complex environment with filamentary structures, which is illustrated in Figure~\ref{figset:3d_JADES-GN-1}.

\setcounter{figure}{3}

\begin{figure*}
    \centering
    \includegraphics[width=0.7\linewidth]{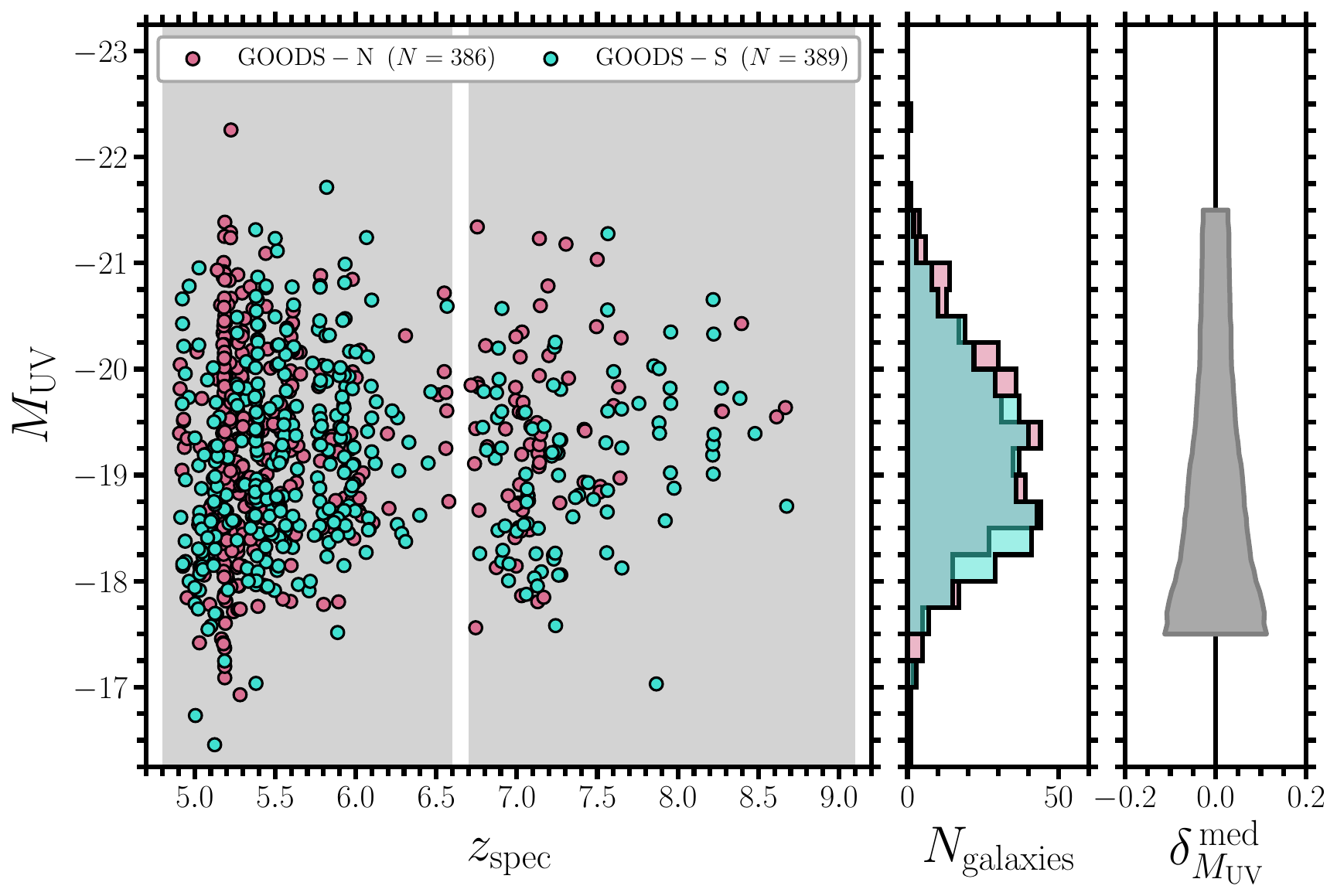}
    \caption{Rest-UV magnitudes versus spectroscopic redshifts for the final spectroscopic sample described in Section~\ref{SectionThreeTwo}. The pink (blue) points represent the galaxies from GOODS-N (GOODS-S). Similarly, histograms showing the distribution of UV magnitudes in the twin axes closest to the right. The pink (blue) shaded regions represent the galaxies from GOODS-N (GOODS-S). The additional depth of JADES in GOODS-S when compared to GOODS-N has resulted in an increased number of very UV-faint galaxies (i.e., $M_{\mathrm{UV}} > -18.5$). The twin axis furthest to the right displays the median uncertainty on the UV magnitudes as a function of UV magnitude. \label{fig:Figure_04}}
\end{figure*}

JADES$-$GS$-$OD$-7.061$ is an overdensity in GOODS-S with $5$ galaxies. The average overdensity value is $\delta_{\mathrm{gal}} = 3.866$ across an inner 68th percentile range of $3.537$ and $4.218$. The angular (physical) diameter is $2.5$ arcmin ($6.5$ cMpc). JADES$-$GS$-$OD$-7.231$ contains $5$ galaxies. The average overdensity value is $\delta_{\mathrm{gal}} = 6.805$ across a similar range of $5.005$ and $8.369$. The angular (physical) diameter is $2.0$ arcmin ($5.2$ cMpc). JADES$-$GS$-$OD$-7.265$ contains $7$ galaxies. The average overdensity value is $\delta_{\mathrm{gal}} = 6.631$ across a similar range of $5.500$ and $7.642$. The angular (physical) diameter is $1.7$ arcmin ($4.4$ cMpc). JADES$-$GS$-$OD$-7.561$ contains $7$ galaxies. The average overdensity value is $\delta_{\mathrm{gal}} = 6.631$ across a similar range of $6.395$ and $6.649$. The angular (physical) diameter is $2.7$ arcmin ($7.0$ cMpc). These overdensities were discussed previously in \citet[][]{Endsley:2023b} as a very dense concentration of $z \approx 7.0-7.6$ galaxies. Furthermore, the extreme LAE discovered by \citet[][JADES$-$GS$-z7-$LA]{Saxena:2023} was later found to coincide with a large-scale galaxy overdensity at $z \approx 7.3$, potentially corresponding to an ionized bubble with a size of $R_{\mathrm{ion}} = 19$ cMpc \citep[][]{Witstok:2023}. However, this overdensity does not contain JADES$-$GS$-z7-$LA, since it is too faint to be detected by the JWST/NIRCam WFSS from FRESCO. 

\begin{figure*}
    \centering
    \includegraphics[width=1.0\linewidth]{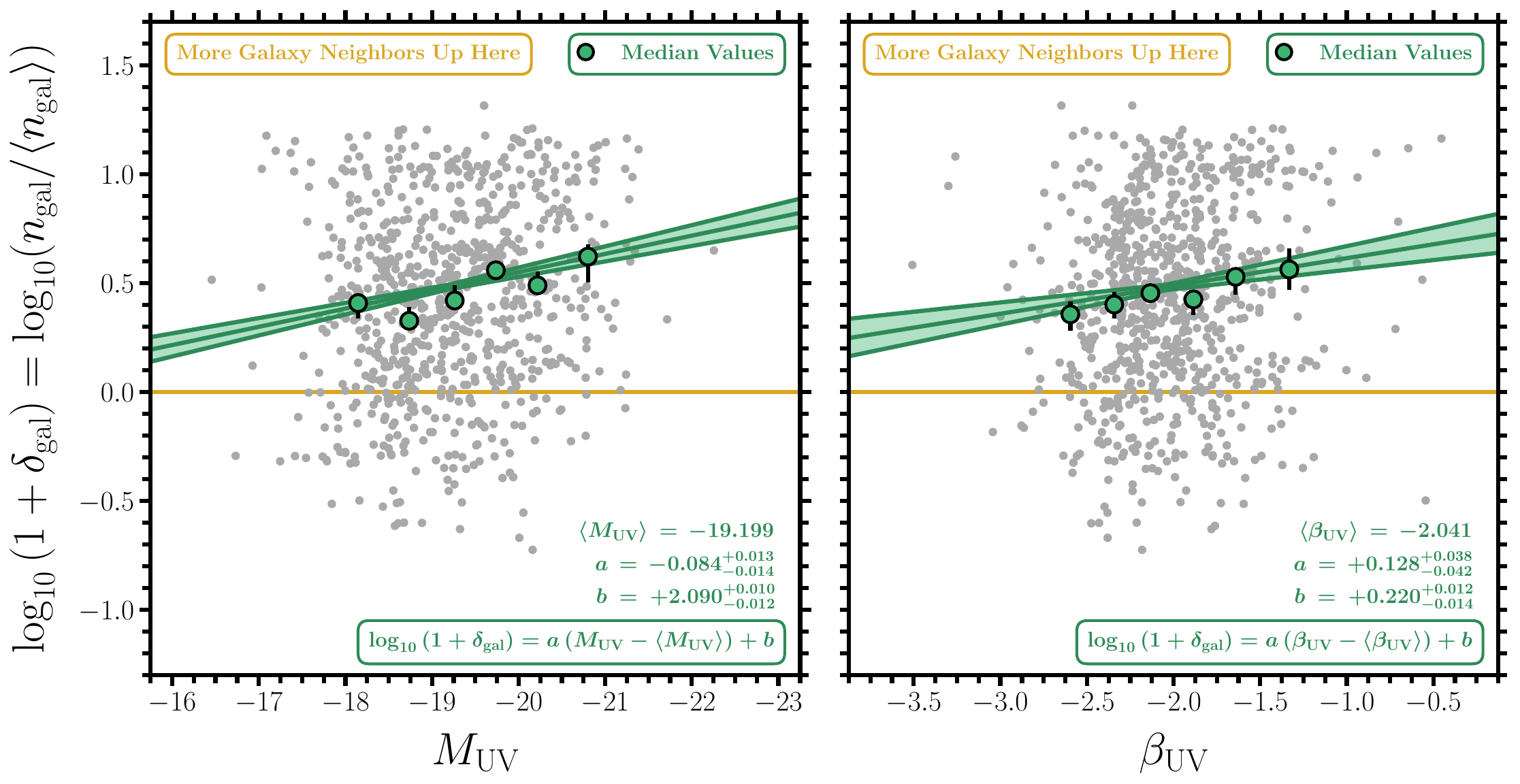}
    \caption{\textit{Left panel}: Galaxy overdensity values versus rest-UV magnitudes for the final spectroscopic sample described in Section~\ref{SectionThreeTwo}. \textit{Right panel}: Galaxy overdensity values versus rest-UV continuum slopes for the final spectroscopic sample. \textit{Both panels}: The grey points represent individual galaxies from the sample, the green points represent median values in bins of UV magnitude or UV continuum slope, and the best-fit linear relation to the median values are shown by the green lines and shaded regions. Best-fit slopes and normalizations are provided in the lower right corners. These results indicate that galaxy overdensity values are significantly correlated with both UV magnitudes (at roughly $6.2\sigma$) and UV continuum slopes (at roughly $3.2\sigma$), suggesting the UV-brightest and UV-reddest objects are typically surrounded by more galaxy neighbors. These correlations provide evidence for accelerated galaxy evolution within overdense environments. \label{fig:Figure_05}}
\end{figure*}

JADES$-$GS$-$OD$-7.954$ is an overdensity in GOODS-S with $4$ galaxies. The average spectroscopic redshift is $z_{\mathrm{\,spec}} = 7.954$ across an inner 68th percentile range of $7.952$ and $7.955$, while the average overdensity value is $\delta_{\mathrm{gal}} = 5.495$ across a similar range of $5.343$ and $5.641$. The angular (physical) diameter is $1.5$ arcmin ($3.9$ cMpc). JADES$-$GS$-$OD$-8.220$ contains $6$ galaxies. The average spectroscopic redshift is $z_{\mathrm{\,spec}} = 8.220$ across an inner 68th percentile range of $8.217$ and $8.222$, while the average overdensity value is $\delta_{\mathrm{gal}} = 10.735$ across a similar range of $10.058$ and $11.225$, making it the most significant overdensity in terms of average overdensity value in either field. The angular (physical) diameter is $2.2$ arcmin ($5.8$ cMpc). These overdensities have not been discussed previously in the literature, although two of the constituent galaxies have been studied by \citet[][]{Laporte:2023} with JWST/NIRCam imaging from JEMS and JWST/NIRCam WFSS from FRESCO. Furthermore, these are the two highest redshift spectroscopically confirmed galaxy overdensities (or protocluster candidates) to date, more distant than the protocluster candidate A$2744-z7$p$9$OD at $\left< z_{\mathrm{\,spec}} \right> = 7.881$ \citep[][]{Morishita:2023, Hashimoto:2023}.

\subsection{Properties of the Rest-UV Photometry}
\label{SectionThreeFour}

Following the methodology outlined in \citet[][]{Topping:2023}, we measure the rest-UV magnitudes ($M_{\mathrm{UV}}$) and continuum slopes ($\beta_{\mathrm{UV}}$) for each of the galaxies that are part of the final spectroscopic sample described in Section~\ref{SectionThreeTwo}. UV luminosities are derived as $\nu L_{\nu}$ at $\lambda_{\mathrm{rest}} = 1500$\AA\ while UV continuum slopes are derived from fitting a simple power law \citep[$f_{\lambda} \propto \lambda^{\beta_{\mathrm{UV}}}$][]{Calzetti:1994} to the rest-UV photometry. We adopt a redshift-dependent set of filters that are chosen to avoid contamination from $\mathrm{Ly}\alpha$ emission to perform these derivations. For objects in our sample at $z_{\,\mathrm{spec}} < 5.4$, the filter set includes F090W, F115W, and F150W. For objects in our sample at $5.4 < z_{\,\mathrm{spec}} < 7.4$, the filter set includes F115W, F150W, and F200W. For objects in our sample at $z_{\,\mathrm{spec}} > 7.4$, the filter set includes F150W, F200W, and F277W. \par

Figure~\ref{fig:Figure_04} shows the distribution of rest-UV magnitudes versus spectroscopic redshifts for the final spectroscopic sample, where the pink points represent the galaxies from GOODS-N and the blue points represent the galaxies from GOODS-S. The twin axes closest to the right include histograms showing the distribution of UV magnitudes for the two fields, assuming a similar color scheme to the primary axis. Since the GOODS-S region of JADES is marginally deeper than the GOODS-N region, there is an increased number (roughly $50\%$ more) of very UV-faint galaxies (i.e., $M_{\mathrm{UV}} > -18.5$) within GOODS-S. To assess the typical uncertainties associated with these values, the twin axis furthest to the right displays the median uncertainty on the UV magnitudes, as a function of UV magnitude. Our sample spans a large range of rest-UV magnitudes, from $M_{\mathrm{UV}} = -16.5$ at the faint end (JADES$-$GS$+$53.10188$-$27.81095) all the way out to $M_{\mathrm{UV}} = -22.3$ at the bright end (JADES$-$GN$+$189.24919$+$62.20523).

Figure~\ref{fig:Figure_05} shows the distribution of galaxy overdensity values versus rest-UV magnitudes in the left axis and galaxy overdensity values versus rest-UV continuum slopes in the right axis for the final spectroscopic sample, where the grey points represent individual galaxies from both GOODS-N and GOODS-S. These UV magnitudes have an inner 68th percentile range in $M_{\mathrm{UV}}$ of $-18.4$ and $-20.1$, with a median UV magnitude of $M_{\mathrm{UV}} = -19.2$, which is broadly consistent (but a little brighter) than the values derived by \citet[][]{Topping:2023} for a similar sample of galaxies that were photometrically selected down to a fainter UV luminosity threshold. These UV continuum slopes have an inner 68th percentile range in $\beta_{\mathrm{UV}}$ of $-2.4$ and $-1.7$, with a median UV continuum slope of $\beta_{\mathrm{UV}} = -2.0$, which is broadly consistent (but a little redder) than the analogous values derived by \citet[][]{Topping:2023}. 

Qualitatively, the distribution of galaxy overdensity values appears to shift lower at fainter UV magnitudes and bluer UV continuum slopes. We explore this apparent trend by assigning objects in our final spectroscopic sample to bins in both $M_{\mathrm{UV}}$ and $\beta_{\mathrm{UV}}$, which are presented in Table~\ref{tab:bins}. As a reminder, 3d galaxy overdensity values are defined to be $\delta_{\mathrm{gal}} = n_{\mathrm{gal}} / \langle n_{\mathrm{gal}} \rangle - 1$, following the methodology of \citet[][]{Chiang:2013}. The UV-brightest objects at these redshifts ($M_{\mathrm{UV}} < -20.5$ at $4.9 < z_{\,\mathrm{spec}} < 8.9$) have median galaxy overdensity values of $\delta_{\mathrm{gal}} = 3.14^{+0.10}_{-0.10}$ (corresponding to number densities that are roughly $4.1$ times that of the ensemble average number density of galaxies), while the UV-faintest objects ($M_{\mathrm{UV}} > -18.5$) have median galaxy overdensity values of $\delta_{\mathrm{gal}} = 1.52^{+0.13}_{-0.16}$ (which is roughly $2.5$ times more dense than average). On the other hand, the UV-reddest objects ($\beta_{\mathrm{UV}} > -1.5$) have median galaxy overdensity values of $\delta_{\mathrm{gal}} = 2.52^{+0.45}_{-0.32}$ (which is roughly $3.5$ times more dense than average), while the UV-bluest objects ($\beta_{\mathrm{UV}} < -2.5$) have median galaxy overdensity values of $\delta_{\mathrm{gal}} = 1.28^{+0.24}_{-0.19}$ (which is roughly $2.3$ times more dense than average). 

It is interesting to note that nearly all of the galaxies in the final spectroscopic sample (roughly $83$\%) have galaxy overdensity values greater than zero, implying that they live in environments that are more dense than the ensemble average number density of galaxies at those redshifts. \textcolor{black}{This is the expected behavior when working with galaxies that are inherently biased, such as those from the line flux limited sample that is used throughout this work: populations of galaxies with stronger bias are more clustered than those with weaker bias \citep[e.g.,][]{Chiang:2013}. As a sanity check, we compare this result with predictions from the First Light And Reionisation Epoch Simulations \citep[FLARES;][]{Lovell:2021}, which are a set of cosmological zoom-in simulations using the EAGLE model \citep[][]{Crain:2015, Schaye:2015}. With these simulations, \citet[][]{Lovell:2021} explored the environmental dependence of high-redshift galaxy evolution, finding that galaxies predominantly live in overdense environments across the full range of stellar masses and SFRs at $5 < z < 9$. Their results show that roughly $80-90\%$ of galaxies have galaxy overdensity values greater than zero, consistent with what we measure for the final spectroscopic sample.}

We quantify these observed trends of increasing galaxy overdensity values at brighter UV magnitudes $M_{\mathrm{UV}}$ and redder UV continuum slopes $\beta_{\mathrm{UV}}$ by fitting a linear relation to the median overdensity values for the magnitude and continuum slope bins that are presented in Table~\ref{tab:bins}. The assumed functional forms for the linear relations are given in the lower right corners of Figure~\ref{fig:Figure_05}, along with the best-fit parameters and the average values (which are used as the zero-points for the relations). We derive uncertainties on the best-fit relations by adopting a bootstrap Monte-Carlo method, where we first construct mock samples consisting of objects selected randomly from the observed sample with replacement. For each object in each mock sample, we perturb the galaxy overdensity value and either the UV magnitude or the UV continuum slope based on their associated uncertainties. We assign the resulting values to the same magnitude or continuum slope bin, and fit the binned values $1000$ times. The uncertainties on the best-fit relations are then taken to be the inner 68th percentile range based on the resulting distribution of best-fit linear parameters from the Monte Carlo routine.

\begin{table*}
	\caption{A summary of physical quantities in bins of UV magnitude and UV continuum slope, described in Section~\ref{SectionThreeFour}.}
	\label{tab:bins}
	\begin{threeparttable}
        \makebox[\textwidth]{
	\hspace*{-20mm}
	\begin{tabular}{cccccc} 
		\hline
		\hline
		Magnitude or Slope Bin & $N_{\mathrm{galaxies}}$\tnote{a} & $z_{\,\mathrm{spec}}$\tnote{b} & $M_{\mathrm{UV}}$\tnote{c} & $\beta_{\mathrm{UV}}$\tnote{d} & $\delta_{\mathrm{gal}}$\tnote{e} \\
		\hline
            $-15.5 > M_{\mathrm{UV}} > -18.5$ & $169$ & $5.380^{+0.013}_{-0.083}$ & $-18.155^{+0.021}_{-0.019}$ & $-2.011^{+0.031}_{-0.031}$ & $1.524^{+0.131}_{-0.158}$ \\
            $-18.5 > M_{\mathrm{UV}} > -19.0$ & $162$ & $5.435^{+0.077}_{-0.026}$ & $-18.735^{+0.014}_{-0.013}$ & $-2.152^{+0.029}_{-0.030}$ & $1.192^{+0.118}_{-0.093}$ \\
            $-19.0 > M_{\mathrm{UV}} > -19.5$ & $156$ & $5.602^{+0.073}_{-0.055}$ & $-19.259^{+0.012}_{-0.011}$ & $-2.097^{+0.023}_{-0.023}$ & $1.613^{+0.152}_{-0.102}$ \\
            $-19.5 > M_{\mathrm{UV}} > -20.0$ & $133$ & $5.588^{+0.186}_{-0.045}$ & $-19.741^{+0.013}_{-0.014}$ & $-2.048^{+0.024}_{-0.023}$ & $2.579^{+0.100}_{-0.106}$ \\
            $-20.0 > M_{\mathrm{UV}} > -20.5$ & $88$ & $5.510^{+0.062}_{-0.085}$ & $-20.216^{+0.010}_{-0.012}$ & $-2.028^{+0.026}_{-0.024}$ & $2.129^{+0.125}_{-0.100}$ \\
            $-20.5 > M_{\mathrm{UV}} > -23.5$ & $62$ & $5.391^{+0.052}_{-0.020}$ & $-20.811^{+0.014}_{-0.013}$ & $-1.824^{+0.026}_{-0.023}$ & $3.137^{+0.097}_{-0.095}$ \\
            \hline
            $-4.00 < \beta_{\mathrm{UV}} < -2.50$ & $66$ & $5.649^{+0.260}_{-0.074}$ & $-18.851^{+0.057}_{-0.046}$ & $-2.650^{+0.020}_{-0.022}$ & $1.279^{+0.238}_{-0.192}$ \\
            $-2.50 < \beta_{\mathrm{UV}} < -2.25$ & $147$ & $5.543^{+0.056}_{-0.037}$ & $-19.181^{+0.087}_{-0.069}$ & $-2.353^{+0.009}_{-0.009}$ & $1.488^{+0.195}_{-0.169}$ \\
            $-2.25 < \beta_{\mathrm{UV}} < -2.00$ & $208$ & $5.541^{+0.031}_{-0.029}$ & $-19.233^{+0.040}_{-0.039}$ & $-2.125^{+0.009}_{-0.009}$ & $1.701^{+0.182}_{-0.166}$ \\
            $-2.00 < \beta_{\mathrm{UV}} < -1.75$ & $182$ & $5.522^{+0.063}_{-0.079}$ & $-19.268^{+0.077}_{-0.088}$ & $-1.885^{+0.008}_{-0.010}$ & $1.781^{+0.230}_{-0.166}$ \\
            $-1.75 < \beta_{\mathrm{UV}} < -1.50$ & $99$ & $5.298^{+0.084}_{-0.013}$ & $-19.355^{+0.117}_{-0.094}$ & $-1.641^{+0.012}_{-0.011}$ & $2.246^{+0.209}_{-0.370}$ \\
            $-1.50 < \beta_{\mathrm{UV}} < -0.00$ & $68$ & $5.340^{+0.037}_{-0.044}$ & $-19.305^{+0.103}_{-0.077}$ & $-1.320^{+0.022}_{-0.021}$ & $2.517^{+0.454}_{-0.321}$ \\
            \hline
	\end{tabular}
        }
	\begin{tablenotes}
	    \footnotesize
            \item \textbf{Notes.}
            \item[a] Total number of galaxies.
            \item[b] Median and 68\% confidence interval of the spectroscopic redshifts.
            \item[c] Median and 68\% confidence interval of the rest-UV absolute magnitudes.
            \item[d] Median and 68\% confidence interval of the rest-UV continuum slopes.
            \item[e] Median and 68\% confidence interval of the galaxy overdensity values.
        \end{tablenotes}
	\end{threeparttable}
\end{table*} 

The results of fitting these linear relations reveal significant correlations between galaxy overdensity values $\delta_{\mathrm{gal}}$ and UV magnitude $M_{\mathrm{UV}}$ as well as UV continuum slope $\beta_{\mathrm{UV}}$. For the relation with UV magnitude, we find a best-fit slope of $a = -0.084^{+0.013}_{-0.014}$ and normalization of $b = 2.090^{+0.010}_{-0.012}$, which are defined at the average value of $M_{\mathrm{UV}} = -19.199$ and are inconsistent with no correlation between $\delta_{\mathrm{gal}}$ and $M_{\mathrm{UV}}$ at $> 6\sigma$. For the relation with UV continuum slope, we find a best-fit slope of $a = 0.128^{+0.038}_{-0.042}$ and normalization of $b = 0.220^{+0.012}_{-0.014}$, which are defined at the average value of $\beta_{\mathrm{UV}} = -2.041$ and are inconsistent with no correlation between $\delta_{\mathrm{gal}}$ and $\beta_{\mathrm{UV}}$ at $> 3\sigma$. Furthermore, we conduct Spearman correlation tests on the individual objects in the final spectroscopic sample, which reveal significant correlations between $\delta_{\mathrm{gal}}$ and $M_{\mathrm{UV}}$ (described by a correlation coefficient of $\rho = -0.159$ and a probability of being drawn from an uncorrelated distribution of $1.0 \times 10^{-5}$) as well as $\beta_{\mathrm{UV}}$ (described by a correlation coefficient of $\rho = 0.123$ and a probability of being drawn from an uncorrelated distribution of $6.4 \times 10^{-4}$).

These significant correlations can be interpreted as evidence for accelerated galaxy evolution within overdense environments, since galaxies with brighter UV absolute magnitudes $M_{\mathrm{UV}}$ and redder UV continuum slopes $\beta_{\mathrm{UV}}$ typically have more galaxy neighbors than their fainter and bluer counterparts. We should note that the UV continuum slope exhibits a well-known dependence on the UV absolute magnitude, and that the slope of this correlation is seemingly independent of redshift \citep[e.g.,][]{Bouwens:2009, Bouwens:2012, Bouwens:2014, Topping:2023}. This may suggest that the correlation between galaxy overdensity value and UV continuum slope is second-order and the result of the correlation between UV absolute magnitude and UV continuum slope. Future work will focus on a detailed comparison of various galaxy properties as a function of local environment.

\subsection{Properties of the Stellar Populations}
\label{SectionThreeFive}

Following the methodology outlined in \citet[][]{Tacchella:2022b} and \citet[][]{Helton:2023}, we utilize the SED fitting code \texttt{Prospector} \citep[v1.1.0;][]{Johnson:2021} to infer the stellar populations for each of the galaxies that are part of the final spectroscopic sample described in Section~\ref{SectionThreeTwo}. Fits are performed on the rescaled Kron small photometry described in Section~\ref{SectionTwoOne}, while the redshift is fixed at the spectroscopic redshift. \texttt{Prospector} uses a Bayesian inference framework and we choose to sample posterior distributions with the dynamic nested sampling code \texttt{dynesty} \citep[v1.2.3;][]{Speagle:2020}. 

\begin{figure*}
    \centering
    \includegraphics[width=1.0\linewidth]{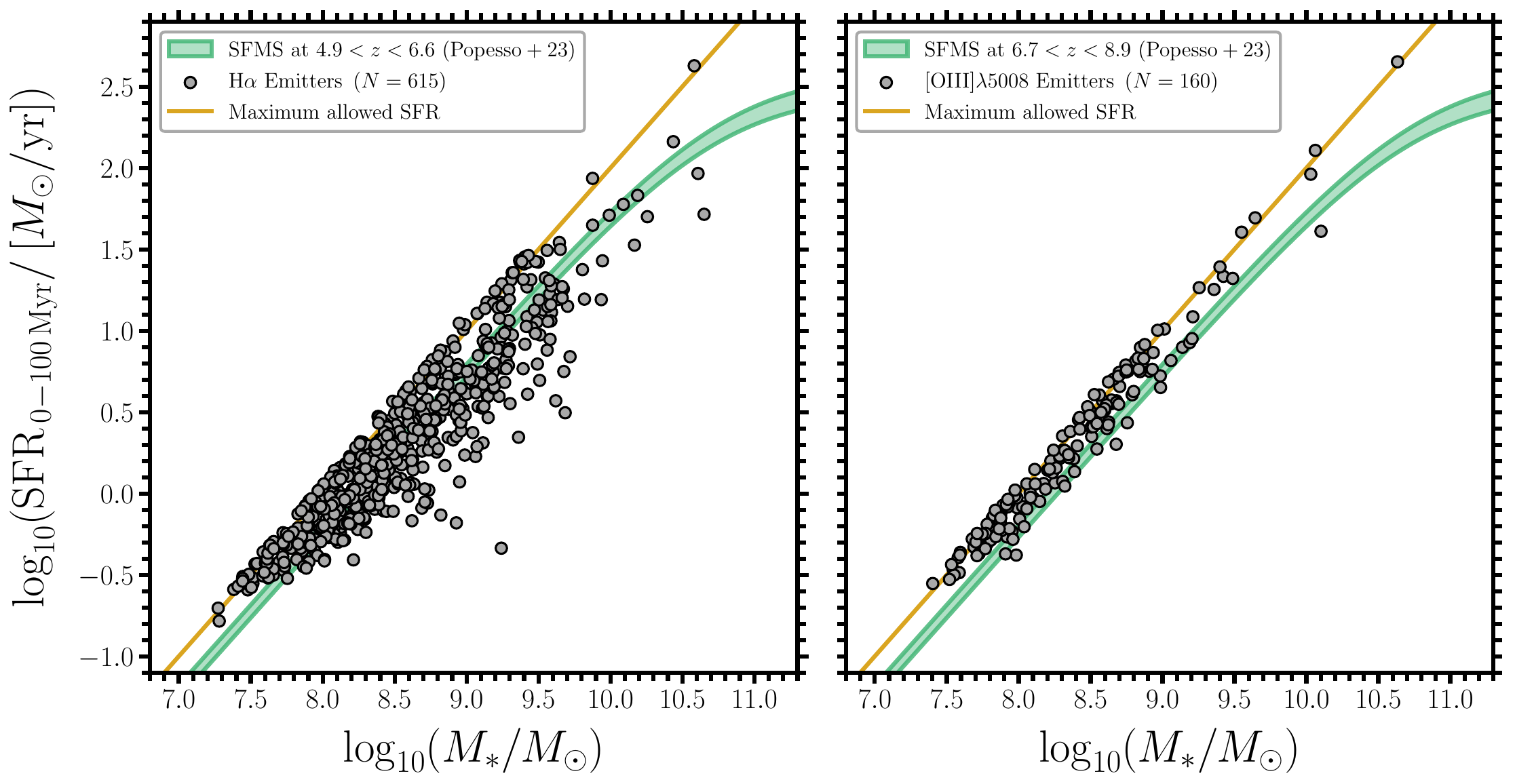}
    \caption{\textit{Left panel}: Star-forming main sequence for the final spectroscopic sample of $\mathrm{H}\alpha$ emitters described in Section~\ref{SectionThreeTwo}. \textit{Right panel}: Star-forming main sequence for the final spectroscopic sample of $\mathrm{[OIII]}\lambda5008$ emitters described in Section~\ref{SectionThreeTwo}. \textit{Both panels}: The stellar masses and SFRs reported here are derived from the \texttt{Prospector} fits, where the SFRs are averaged over the last $100\ \mathrm{Myr}$ of lookback time. The grey points represent individual galaxies from the sample. The empirical star-forming main sequence at $4.9 < z < 6.6$ and $6.7 < z < 8.9$, as derived by \citet{Popesso:2023}, is given by the green shaded region. The maximum allowed SFR assuming all of the stellar mass was formed in the last $100\ \mathrm{Myr}$ of lookback time is given by the solid gold line. The vast majority of these objects agree with the empirically derived star-forming main sequence within $1\sigma$. \label{fig:Figure_06}}
\end{figure*}

We use the Flexible Stellar Population Synthesis \citep[\texttt{FSPS};][]{Conroy:2009, Conroy:2010} code, which is accessed through the \texttt{python-FSPS} \citep{Foreman-Mackey:2014} bindings. We assume the MIST isochrones and stellar tracks \citep{Dotter:2016, Choi:2016} which make use of the MESA stellar evolution package \citep{Paxton:2011, Paxton:2013, Paxton:2015, Paxton:2018}. We further assume the MILES stellar spectral library \citep{Falcon-Barroso:2011, Vazdekis:2015} and adopt a \citet{Chabrier:2003} initial mass function (IMF). Absorption by the intergalactic medium (IGM) is modeled after \citet{Madau:1995}, where the overall scaling of the IGM attenuation curve is set to be a free parameter. Dust attenuation is modeled using a two-component dust attenuation model \citep{Charlot:2000} with a flexible attenuation curve where the slope is tied to the strength of the UV bump \citep{Kriek:2013}. Nebular emission (both from emission lines and continuum) is self-consistently modeled with the spectral synthesis code \texttt{Cloudy} \citep{Ferland:2013, Byler:2017}. 

A non-parametric SFH is used with the ``continuity'' prior \citep{Tacchella:2022b}, which allows for large flexibility in the shape of the SFH since they are not parametric functions of time. We assume that the SFH can be described by $N_{\mathrm{SFR}}$ distinct time bins of constant star formation. The time bins are specified in units of lookback time and the number of distinct bins is set to be $N_{\mathrm{SFR}} = 6$. The first two bins are fixed at $0-30\ \mathrm{Myr}$ and $30-100\ \mathrm{Myr}$ while the last bin is fixed between $0.15\,t_{\mathrm{univ}}$ and $t_{\mathrm{univ}}$, where $t_{\mathrm{univ}}$ is the age of the Universe at the galaxy's spectroscopic redshift, measured with respect to the formation redshift $z_{\,\mathrm{form}} = 20$. The rest of the bins are spaced equally in logarithmic time between $100\,\mathrm{Myr}$ and $0.85\,t_{\mathrm{univ}}$. Both the total stellar mass and the ratios of SFRs in adjacent time bins are set to be free parameters.

Figure~\ref{fig:Figure_06} shows the star-forming main sequence for the final spectroscopic sample identified in Section~\ref{SectionThreeTwo}. The reported stellar masses and SFRs are derived from the \texttt{Prospector} fits using the non-parametric SFH with the ``continuity'' prior, where the SFRs are averaged over the last $100\ \mathrm{Myr}$ of lookback time. The empirical star-forming main sequence at $4.9 < z < 8.9$ derived by \citet{Popesso:2023} is given by the green shaded region. Additionally, the maximum allowed SFR assuming all of the stellar mass was formed in the last $100\ \mathrm{Myr}$ of lookback time is given by the solid gold line. We note that the empirical star-forming main sequence used here is only calibrated at redshifts of $0 < z < 6$ and stellar masses of $10^{8.5} M_{\odot} < M_{\ast} < 10^{11.5} M_{\odot}$, which does not cover the full range of values for the final spectroscopic sample of galaxies considered here.

We find that the vast majority of the objects in our sample agree with the empirically derived star-forming main sequence within $1\sigma$ \citep{Popesso:2023}, despite our sample being biased as a result of the requirement to detect either $\mathrm{H}\alpha$ or $\mathrm{[OIII]}\lambda5008$ at greater than $3\sigma$. For the galaxies with $\mathrm{H}\alpha$ detections at $4.9 < z_{\,\mathrm{spec}} < 6.6$, we could alternatively derive SFRs using the measured $\mathrm{H}\alpha$ emission line flux along with the conversion factor from \citet{Kennicutt:2012}. These objects would have their SFRs increased by $\Delta \sim 0.5\,\mathrm{dex}$ if the $\mathrm{H}\alpha$-based SFRs were adopted rather than those derived from the \texttt{Prospector} fits. We note that this is likely because the canonical hydrogen ionizing photon production efficiency ($\xi_\mathrm{ion} \propto L_{\mathrm{H}\alpha}/L_{\mathrm{UV}}$) used in \citet{Kennicutt:2012} for galaxies in the local Universe with solar metallicity is only $\xi_\mathrm{ion} \sim 10^{25.1}\,\mathrm{Hz}/\mathrm{erg}$, which is lower than measurements at comparable redshifts by $\Delta \xi_\mathrm{ion} \sim 0.5\,\mathrm{dex}$ \citep[e.g.,][]{Bouwens:2016, Ning:2023, Sun:2023a, Simmonds:2023}. For the galaxies with $\mathrm{[OIII]}\lambda5008$ detections at $6.7 < z_{\,\mathrm{spec}} < 8.9$, we could alternatively derive SFRs using the measured $\mathrm{[OIII]}\lambda5008$ emission line flux along with the conversion factor from \citet{Villa-Velez:2021}. These objects would similarly have their SFRs increased by $\Delta \sim 0.5\,\mathrm{dex}$ if the $\mathrm{[OIII]}\lambda5008$-based SFRs were adopted rather than those derived from the \texttt{Prospector} fits, which is also likely caused by the redshift evolution of $\xi_\mathrm{ion}$ as shown with recent literature \citep[][Sun et al., in preparation]{Simmonds:2023, Tang:2023}. The use of UV-based SFRs derived using the conversion factor from \citet{Kennicutt:2012} does not change the distribution of our final spectroscopic sample on Figure~\ref{fig:Figure_06}, which is expected, since the \texttt{Prospector}-based SFRs largely rely on the rest-UV photometry.

Within the final spectroscopic sample described in Section~\ref{SectionThreeTwo}, there are $N = 615$ galaxies with $\mathrm{H}\alpha$ detections across the redshift range $4.9 < z_{\,\mathrm{spec}} < 6.6$, which span over three orders of magnitude in inferred stellar mass from $M_{\ast} = 10^{7.3}\,M_{\odot}$ up to $M_{\ast} = 10^{10.7}\,M_{\odot}$. The least massive galaxy in this lower-redshift sample (JADES$-$GS$+$53.16514$-$27.85392) is also one of the UV-faintest in the sample with a magnitude of $M_{\mathrm{UV}} = -17.6$ at $z_{\,\mathrm{spec}} = 5.101$ while the most massive galaxy (JADES$-$GN$+$189.11308$+$62.29239) is another one of the UV-faintest in the sample with a continuum slope of $M_{\mathrm{UV}} = -18.0$ at $z_{\,\mathrm{spec}} = 5.270$. Additionally, there are $N = 160$ galaxies with $\mathrm{[OIII]}\lambda5008$ detections across the redshift range $6.7 < z_{\,\mathrm{spec}} < 8.9$, spanning over three orders of magnitude in inferred stellar mass from $M_{\ast} = 10^{7.4}\,M_{\odot}$ up to $M_{\ast} = 10^{10.7}\,M_{\odot}$. The least massive galaxy in this higher-redshift sample (JADES$-$GS$+$53.16242$-$27.75070) is another one of the UV-faintest in the sample with a magnitude of $M_{\mathrm{UV}} = -18.5$ at $z_{\,\mathrm{spec}} = 7.044$ while the most massive galaxy (JADES$-$GS$+$53.16138$-$27.73767) is one of the UV-reddest in the sample with a continuum slope of $\beta_{\mathrm{UV}} = -1.4$ at $z_{\,\mathrm{spec}} = 6.812$. 
We also note that both of the two most massive sources have heavily reddened SEDs ($A_{V} \approx 3\,\mathrm{mag}$) and point-like morphologies, and therefore we cannot rule out the existence of broad-line AGN as those seen with recent JWST studies \citep[e.g.,][]{Matthee:2023, Maiolino:2023, Greene:2023}, which could be fainter than the current grism detection limit. If so, the stellar masses derived from the \texttt{Prospector} fits could be significantly overestimated for these two sources.

A large fraction of the galaxies within the final spectroscopic sample agree with the maximum allowed SFR line within $1\sigma$, which implies that these galaxies assembled nearly all of their stellar mass in the last $100\ \mathrm{Myr}$ of lookback time, suggesting a strong recent burst of star formation. Such bursty SFHs are seemingly common in galaxies at these redshifts, as suggested by recent observations with JWST \citep[e.g.,][]{Endsley:2023a, Endsley:2023b, Dome:2023, Looser:2023a, Looser:2023b, Simmonds:2023, Strait:2023, Tacchella:2023b}. There is one interesting outlier well below the empirical star-forming main sequence in the left panel of Figure~\ref{fig:Figure_06} (JADES$-$GN$+$189.23669$+$62.20247 at $z_{\,\mathrm{spec}} = 5.200$). With an inferred stellar mass of $M_{\ast} = 10^{9.2}\,M_{\odot}$, the observed photometry suggests a strong Balmer break, implying a mass-weighted age of $t_{\mathrm{MW}} = 390\,\mathrm{Myr}$. This galaxy is part of the largest overdensity identified here in terms of total number of galaxies and inferred halo mass (JADES$-$GN$-$OD$-5.191$), with an angular (physical) separation of $0.6$ arcmin ($1.5$ cMpc) from the well-known sub-millimeter galaxy HDF850.1. We do not expect to find many galaxies with strong Balmer breaks, as a result of our selection.

\vspace{-2dd}
\subsection{Inferring Dark Matter Halo Masses}
\label{SectionThreeSix}

The stellar-to-halo abundance matching relation from \citet{Behroozi:2013} is typically used to convert stellar masses into halo masses, and therefore to derive estimates of the dark matter halo mass for individual overdensities and protoclusters in the early Universe \citep[e.g.,][]{Long:2020, Helton:2023}. Uncertainties on these estimates are calculated by adopting the mean stellar-to-halo abundance matching relation from \citet{Behroozi:2013} and propagating the uncertainties on the stellar masses for each individual galaxy. However, it is unclear how the assumed relation changes when working with biased populations of galaxies, such as the line flux limited sample that is used throughout this work. For this reason, it is important for us to derive our own stellar-to-halo abundance matching relations, from simulated samples of galaxies that match the final spectroscopic sample described in Section~\ref{SectionThreeTwo} in terms of selection.

We make use of the mock catalogs and lightcones produced for each of the five CANDELS fields \citep{Grogin:2011} in the public data release (DR1) of the semi-empirical UniverseMachine model from \citet{Behroozi:2019}. The model obtained their dark matter halo properties and assembly histories from the Bolshoi-Planck dark matter halo simulations \citep[][]{Klypin:2016, Rodriguez-Puebla:2016}. These simulations followed a periodic, comoving volume with side lengths of $250 h^{-1}$ Mpc containing $2048^{3}$ dark matter particles with a mass resolution of $1.6 \times 10^{8} h^{-1}\, M_{\odot}$. The assumed cosmology for these simulations is the standard flat $\Lambda$CDM cosmology from Planck15, which is fully consistent with the Planck18 cosmology assumed here. For more information about the mock catalogs and lightcones produced by the semi-empirical UniverseMachine model, we refer the reader to \citet{Behroozi:2019}. % The \texttt{Rockstar} \citep{Behroozi:2013b} and \texttt{ConsistentTrees} \citep{Behroozi:2013d} codes are used for halo finding and merger tree construction, respectively. Galaxy populations are constructed within these simulated dark matter halos to fit recent constraints from observations, including: stellar mass functions, cosmic and specific star formation rates, quenched fractions, UV luminosity functions, UV to stellar mass relations, infrared excess to UV relations, measurements of galaxy correlation functions, and the environmental dependence of central galaxy quenching. For more information about the mock catalogs and lightcones produced by the semi-empirical UniverseMachine model, we refer the reader to \citet{Behroozi:2019}.

There are 10 different lightcones produced for each of the five CANDELS fields, corresponding to the best-fit model along with 9 randomly drawn models from the posterior distribution of the UniverseMachine DR1. For each of these lightcones, there are 8 different realizations, resulting in a total of 80 unique catalogs that are simulated for each of the fields. Since we are interested in the GOODS-N and GOODS-S fields, there are 160 unique catalogs considered here. These lightcones probe fields-of-views of 544 square arcminutes in GOODS-N and 663 square arcminutes in GOODS-S. We further restrict these lightcones to lie within the total overlapping area between the JADES and FRESCO footprints (which is roughly 35 square arcminutes in GOODS-N, 46 square arcminutes in GOODS-S). 

Finally, we impose cuts on the rest-frame UV magnitudes and SFRs, where both of these are taken to be observed values rather than intrinsic, in order to produce realistic comparison samples of simulated galaxies. The rest-frame UV magnitude cut is a fixed cut that is equal to the faintest UV magnitude in the final spectroscopic sample, $M_{\mathrm{UV}} = -16.5$. The SFR cut is a variable cut that is dependent on the JWST/NIRCam Grism R line flux sensitivity at a given wavelength, and is therefore dependent on the spectroscopic redshift. The line flux sensitivities at $3\sigma$ are used along with SFR conversion factors from \citet{Kennicutt:2012} for $\mathrm{H}\alpha$ and \citet{Villa-Velez:2021} for $\mathrm{[OIII]}\lambda5008$. We further reduce these SFR limits by a factor of three in order to correct for the observed difference between UV-based SFRs and $\mathrm{H}\alpha$-based or $\mathrm{[OIII]}\lambda5008$-based SFRs, as discussed in Section~\ref{SectionThreeFive}. The resulting SFR cut is roughly $1\,M_{\odot}/\mathrm{yr}$ at the blue end ($z_{\,\mathrm{spec}} \approx 5$) of the $\mathrm{H}\alpha$ emitters and roughly $2\,M_{\odot}/\mathrm{yr}$ at the red end ($z_{\,\mathrm{spec}} \approx 6.5$), while the same SFR cut is roughly $2\,M_{\odot}/\mathrm{yr}$ at the blue end ($z_{\,\mathrm{spec}} \approx 7$) of the $\mathrm{[OIII]}\lambda5008$ emitters and roughly $4\,M_{\odot}/\mathrm{yr}$ at the red end ($z_{\,\mathrm{spec}} \approx 9$). We note that the SFR cut is more restrictive when compared to the rest-frame UV magnitude cut, and ultimately is the only relevant cut, since the final spectroscopic sample is line flux limited and therefore implicitly SFR limited. 

\begin{figure*}
    \centering
    \includegraphics[width=1.0\linewidth]{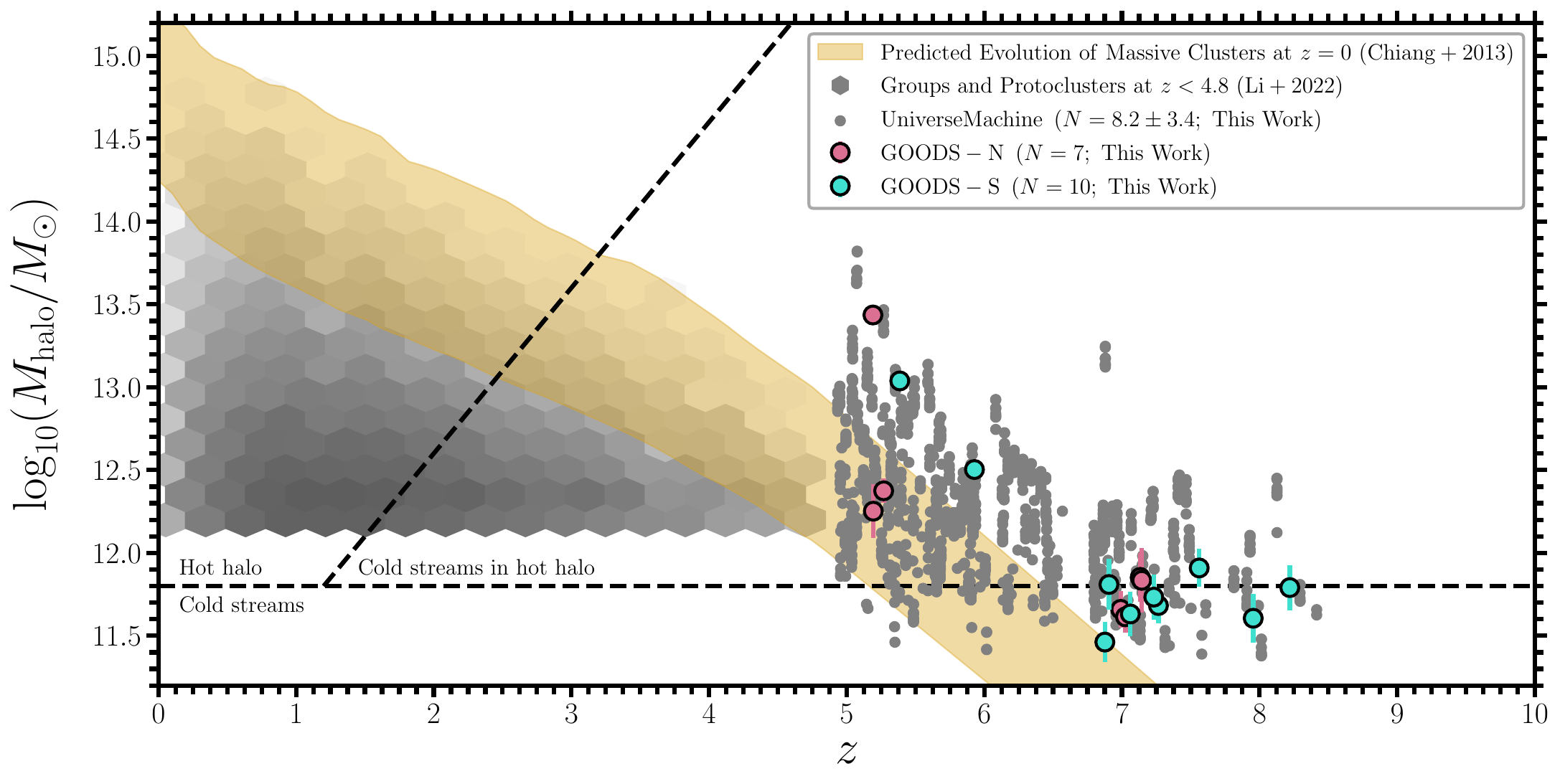}
    \caption{Dark matter halo mass versus redshift for the final sample of galaxy overdensities described in Section~\ref{SectionThreeThree}. The pink (blue) points represent the overdensities from GOODS-N (GOODS-S), while the dark matter halo mass estimates for these structures are described in Section~\ref{SectionThreeSix}. For comparison, groups and protoclusters at $z \lesssim 5$ from \citet{Li:2022} are given in grayscale while overdensities from the UniverseMachine at $z \gtrsim 5$ are given by the grey points. The brown shaded region shows the expected halo mass evolution of massive galaxy clusters at $z = 0$ \citep{Chiang:2013}. The black horizontal dashed line represents the typical threshold for shock stability assuming a spherical infall, below which the flows are predominantly cold and above which a shock-heated ICM is expected \citep{Dekel:2006}. The black diagonal dashed line represents the typical threshold for penetrating cold gas flows. \label{fig:Figure_07}}
\end{figure*}

The aforementioned cuts on the rest-frame UV magnitudes and SFRs are applied to each of the 80 unique catalogs that are simulated for both the GOODS-N and GOODS-S fields, producing final comparison samples of galaxies from the UniverseMachine. Since the goal of these cuts was to produce realistic comparison samples of simulated galaxies, we compare the distributions of various physical parameters ($z_{\,\mathrm{spec}}$, $M_{\mathrm{UV}}$, $M_{\ast}$, and SFR) for these simulated samples with the final spectroscopic sample described in Section~\ref{SectionThreeTwo}. We find that the typical simulated distributions of these four parameters are indistinguishable from the observed distributions in both of the fields considered here, when comparing both the full and inner 68th percentile ranges of values, as well as the medians and means of the values. The only exception occurs at the faint end of the $M_{\mathrm{UV}}$ distribution, where the observed sample extends to fainter magnitudes when compared to the simulated samples. This discrepancy at faint $M_{\mathrm{UV}}$ implies that the observed low-mass galaxies are exhibiting larger scatter in their rest-optical emission line strengths when compared to the simulated low-mass galaxies. Larger scatter in the rest-optical emission line strengths could be the result of bursty SFHs not being properly modeled for the simulated galaxies. Recent observations with JWST have suggested that bursty SFHs are likely common in galaxies at these redshifts \citep[e.g.,][]{Endsley:2023a, Endsley:2023b, Dome:2023, Looser:2023a, Looser:2023b, Simmonds:2023, Strait:2023, Tacchella:2023a}.

Since we have confirmed that the final comparison samples of galaxies from the UniverseMachine matches the final spectroscopic sample described in Section~\ref{SectionThreeTwo} in terms of both selection and galaxy properties, we are now able to derive our own stellar-to-halo abundance matching relations from the simulated samples of galaxies. For each of the galaxies that are part of the final spectroscopic sample, we perturb the stellar mass 100 times based on the associated uncertainties. For each of these perturbations on the stellar mass, we find the simulated galaxy in each of the final comparison samples that has the closest stellar mass and lies at a similar redshift ($\Delta z \leq 0.25$). The stellar-to-halo mass ratios of these simulated galaxies are used to convert the perturbed stellar masses into halo masses, producing a distribution of 8000 halo mass estimates for each of the galaxies that are part of the final spectroscopic sample. The mean and standard deviation of this distribution of values are taken to be the best-fit halo mass estimate and corresponding uncertainty. Dark matter halo mass estimates for each of the galaxy overdensities in the final sample described in Section~\ref{SectionThreeThree} are calculated by summing the halo masses for each of the constituent galaxies. Uncertainties on our estimates are calculated by propagating the uncertainties on the halo masses for each individual galaxy. 

We note that our method for estimating dark matter halo masses does not account for additional members of the overdensities that were not identified and included in the final spectroscopic sample, including objects that fall outside either the JADES or the FRESCO footprints. This is a non-negligible effect since some of the overdensities identified here (e.g., JADES$-$GS$-$OD$-5.386$) fall at the edge of the overlapping area between the JADES and FRESCO footprints ($\Delta \theta \ll 1^{\prime}$). Additionally, since our final spectroscopic sample only includes galaxies with narrow photometric redshift constraints and secure emission line detections, we are likely missing some additional subset of objects with relatively unconstrained photometric redshifts and/or low levels of star formation (e.g., DSFGs, massive quiescent galaxies and/or obscured AGNs). Given that clusters induce earlier quenching for their constituent members, we cannot rule out the existence of a significant population of these kinds of objects. For these reasons, the halo masses quoted here are likely underestimates of the true halo masses associated with the final sample of overdensities described in Section~\ref{SectionThreeThree}.

Figure~\ref{fig:Figure_07} shows the dark matter halo mass distribution as a function of redshift for the final sample of galaxy overdensities described in Section~\ref{SectionThreeThree}, with the pink (blue) points representing the overdensities from GOODS-N (GOODS-S). Groups and protoclusters at $z \lesssim 5$ from \citet{Li:2022} are given in grayscale for comparison, which were selected based on photometric redshifts and have dark matter halo mass estimates that assume the stellar-to-halo abundance matching relation from \citet{Behroozi:2013}. Overdensities from the UniverseMachine at $z \gtrsim 5$ are given by the grey points for an additional point of comparison, which were selected in the same manner as the final sample of overdensities following the procedure described in Section~\ref{SectionThreeThree}. The brown shaded region shows the expected halo mass evolution of massive ($M_{\mathrm{halo}} \gtrsim 10^{14}\, M_{\odot}$) galaxy clusters at $z = 0$ \citep{Chiang:2013} assuming a smooth evolution at $z > 6$. The black dashed line represents the typical threshold for shock stability assuming a spherical infall, below which the flows are predominantly cold and above which a shock-heated ICM is expected \citep{Dekel:2006}. The black diagonal dashed line represents the typical threshold for penetrating cold gas flows.

The theoretical overdensities identified from the UniverseMachine exhibit remarkable similarity with the final sample of overdensities described in Section~\ref{SectionThreeThree}. By looking at the distribution of the number of overdensities identified from each of the 160 unique UniverseMachine lightcones (split evenly between GOODS-N and GOODS-S), we calculate the expected number of observed overdensities by taking the mean and standard deviation of this distribution, which produces an expectation of $N = 8.2 \pm 3.4$ overdensities per field. This expected value is the same for each of the two fields considered here ($N = 8.4 \pm 4.0$ for GOODS-N, $N = 8.2 \pm 3.2$ for GOODS-S) and is consistent with the number of overdensities found within the final sample described in Section~\ref{SectionThreeThree} ($N = 7$ for GOODS-N, $N = 10$ for GOODS-S). Furthermore, the range in values for the other inferred parameters (total number of constituent galaxies, mean of the constituent spectroscopic redshifts, mean of the constituent galaxy overdensity values, and mean of the inferred halo mass) are consistent between the theoretical overdensities identified from the UniverseMachine and the final sample of observed overdensities. % ($4 < N_{\mathrm{galaxies}} < 110$, $4.9 \lesssim \left< z_{\mathrm{\,spec}} \right> \lesssim 8.4$, $3 \lesssim \left< \delta_{\mathrm{gal}} \right> \lesssim 27$, and $11.5 \lesssim \mathrm{log}_{10} \left( M_{\mathrm{halo}}/M_{\odot} \right) \lesssim 13.8$) and the final sample of observed overdensities ($4 < N_{\mathrm{galaxies}} < 103$, $5.1 \lesssim \left< z_{\mathrm{\,spec}} \right> \lesssim 8.3$, $3 \lesssim \left< \delta_{\mathrm{gal}} \right> \lesssim 11$, and $11.5 \lesssim \mathrm{log}_{10} \left( M_{\mathrm{halo}}/M_{\odot} \right) \lesssim 13.4$).

\citet[][]{Chiang:2013} used the Millennium Run dark matter $N$-body simulation \citep[][]{Springel:2006} and a semi-analytic galaxy formation model \citep[][]{Guo:2011} to study the high-redshift progenitors of the most extreme present-day structures and their galaxy contents. This simulation followed a comoving volume with side lengths of $500 h^{-1}$ Mpc containing $2160^{3}$ dark matter particles with a mass resolution of $8.6 \times 10^{8} h^{-1}\, M_{\odot}$. It has already been shown that the cluster abundance, cluster galaxy luminosity function, and galaxy number density profiles in these simulations match observations of galaxy clusters in the local Universe. Using these results, they compiled a sample of $2832$ galaxy clusters which includes $1976$ low-mass ``Fornax-type'' clusters\footnote{``Fornax-type'' refers to relatively low-mass galaxy clusters with halo masses of $M_{\mathrm{halo}} = \left( 1.37-3 \right) \times 10^{14}\, M_{\odot}$ at $z = 0$.}, $797$ intermediate-mass ``Virgo-type'' clusters\footnote{``Virgo-type'' refers to intermediate-mass galaxy clusters with halo masses of $M_{\mathrm{halo}} = \left( 3-10 \right) \times 10^{14}\, M_{\odot}$ at $z = 0$.}, and $59$ high-mass ``Coma-type'' clusters\footnote{``Coma-type'' refers to relatively high-mass galaxy clusters with halo masses of $M_{\mathrm{halo}} > 10^{15}\, M_{\odot}$ at $z = 0$.}. These correspond to number densities of $n = 8.8 \times 10^{-6}\, \mathrm{Mpc}^{-3}$ for all clusters, $n = 6.1 \times 10^{-6}\, \mathrm{Mpc}^{-3}$ for ``Fornax-type'' clusters, $n = 2.5 \times 10^{-6}\, \mathrm{Mpc}^{-3}$ for ``Virgo-type'' clusters, and $n = 1.8 \times 10^{-7}\, \mathrm{Mpc}^{-3}$ for ``Coma-type'' clusters. The total comoving volume sampled by our observations is $V = 7.6 \times 10^{3}\, \mathrm{cMpc}^{3}$, which corresponds to a number density of $n = 2.2 \times 10^{-5}\, \mathrm{cMpc}^{-3}$ for the final sample of overdensities described in Section~\ref{SectionThreeThree}. This number density is largely effected by cosmic variance, but taking it at face value, there seems to be a factor of two to three more overdensities in our observations and in the UniverseMachine than clusters in the simulations used by \citet[][]{Chiang:2013}. 

\textcolor{black}{Throughout this work, we have implicitly assumed that each of the overdensities that are part of the final sample described in Section~\ref{SectionThreeThree} will evolve into distinct galaxy clusters by $z = 0$. However, roughly half of the overdensities in the final sample are spatially and kinematically coincident with one another (e.g., the complex environment with filamentary structures that is illustrated in Figure~\ref{figset:3d_JADES-GN-1}). If we instead assume that these nearby overdensities will merge with one another by $z = 0$, then the total number of galaxy overdensities would be reduced from $N = 17$ to $N = 9$, thereby reducing the number density to $n = 1.2 \times 10^{-5}\, \mathrm{cMpc}^{-3}$ for the final sample of overdensities.} Based on these comparisons, we find it plausible that the galaxy overdensities identified here are likely the progenitors of massive galaxy clusters with halo masses of $M_{\mathrm{halo}} \gtrsim 10^{14}\, M_{\odot}$ at $z = 0$. Future work using results from the UniverseMachine, \textcolor{black}{or some other simulation}, will provide more insight into the potential evolution and eventual fates of our galaxy overdensities.

% ($V = 3.3 \times 10^{3}\, \mathrm{cMpc}^{3}$ in GOODS-N and $V = 4.3 \times 10^{3}\, \mathrm{cMpc}^{3}$ in GOODS-S)
% ($n = 2.1 \times 10^{-5}\, \mathrm{cMpc}^{-3}$ in GOODS-N and $n = 2.3 \times 10^{-5}\, \mathrm{cMpc}^{-3}$ in GOODS-S)

%% Start of section four.
\section{Summary \& Conclusions}
\label{SectionFour}

We have presented a systematic search for high-redshift galaxy overdensities at $4.9 < z_{\,\mathrm{spec}} < 8.9$ in both the GOODS-N and GOODS-S fields using data from the Near Infrared Camera (NIRCam) on the James Webb Space Telescope (JWST). These data consist of JWST/NIRCam imaging from the JWST Advanced Deep Extragalactic Survey \citep[JADES;][]{Eisenstein:2023} and the JWST Extragalactic Medium-band Survey \citep[JEMS;][]{Williams:2023} alongside JWST/NIRCam wide field slitless spectroscopy from the First Reionization Epoch Spectroscopic COmplete Survey \citep[FRESCO;][]{Oesch:2023}. Our findings can be summarized as follows.
\begin{enumerate}
    \item Galaxies were initially selected using HST+JWST photometry spanning $\lambda = 0.4-5.0\ \mu\mathrm{m}$ and the redshifts for roughly a third of these galaxies were subsequently confirmed using slitless spectroscopy over $\lambda = 3.9-5.0\ \mu\mathrm{m}$ via a targeted emission line search for either $\mathrm{H} \alpha$ or $\left[\mathrm{OIII}\right]\lambda5008$ around the best-fit photometric redshift. The final spectroscopic sample of galaxies considered here includes $N = 775$ objects at $4.9 < z_{\,\mathrm{spec}} < 8.9$ ($N = 615$ of these galaxies have $\mathrm{H} \alpha$ detections while $N = 160$ of these galaxies have $\left[\mathrm{OIII}\right]\lambda5008$ detections).
    \item A Friends-of-Friends (FoF) algorithm was used to identify high-redshift galaxy overdensities by iteratively looking for three-dimensional (3d) structural groupings within the final spectroscopic sample. Robust galaxy overdensities were selected from this initial sample by requiring a minimum number of constituent galaxies ($N_{\mathrm{galaxies}} \geq 4$) and an average galaxy overdensity value that is larger than the value taken from \citet{Chiang:2013}, which selects structures within cosmological simulations as protocluster candidates with $80\%$ confidence ($\delta_{\mathrm{gal}} = n_{\mathrm{gal}} / \langle n_{\mathrm{gal}} \rangle - 1 \geq 3.04$). The final sample of galaxy overdensities includes $N = 17$ objects ($N = 7$ of these are in GOODS-N while $N = 10$ of these are in GOODS-S). The two highest redshift spectroscopically confirmed galaxy overdensities (or protocluster candidates) to date are part of this sample at $\left< z_{\mathrm{\,spec}} \right> = 7.954$ and $\left< z_{\mathrm{\,spec}} \right> = 8.222$, representing densities around $\sim 6$ and $\sim 12$ times that of a random volume.
    \item The rest-ultraviolet (UV) magnitudes and continuum slopes for the final spectroscopic sample were inferred from the HST+JWST photometry spanning $\lambda = 0.4-5.0\ \mu\mathrm{m}$ and the spectroscopic redshifts as determined by the targeted line search. We divided these galaxies into bins of UV magnitude and continuum slope in order to explore how these parameters vary as a function of galaxy overdensity value. The results of these tests indicate that galaxy overdensity values are significantly correlated with both UV magnitudes (at roughly $6.2\sigma$) and UV continuum slopes (at roughly $3.2\sigma$), suggesting the UV-brightest and UV-reddest objects are typically surrounded by more galaxy neighbors. These correlations provide evidence for accelerated galaxy evolution within overdense environments.
    \item The total dark matter halo mass associated with each of the galaxy overdensities was estimated ($11.5 \leq \mathrm{log}_{10}\left[M_{\mathrm{halo}}/M_{\odot}\right] \leq 13.4$) using an empirical stellar mass to halo mass relation derived from lightcones produced by the semi-empirical UniverseMachine model from \citet{Behroozi:2019}. The total number of simulated galaxy overdensities selected from these lightcones ($N = 8.2 \pm 3.4$ overdensities per field) are consistent with our observed overdensities. Furthermore, the simulated galaxy overdensities selected from these lightcones exhibit similar physical properties (e.g., redshift, number of constituent galaxies, average galaxy overdensity value, and dark matter halo mass) when compared to our observed overdensities. As a result of our selection criteria, the total dark matter halo mass ranges quoted here are likely an underestimate of the true halo masses. Regardless, these large-scale structures are expected to evolve into massive galaxy clusters with $\mathrm{log}_{10}(M_{\mathrm{halo}}/M_{\odot}) \gtrsim 14$ by $z = 0$.
\end{enumerate}

In this work, we have demonstrated the powerful combination of JWST/NIRCam imaging and slitless spectroscopy by efficiently confirming the redshifts for $N = 775$ objects at $4.9 < z_{\,\mathrm{spec}} < 8.9$, inferring the physical properties of these galaxies, and assessing the large-scale structure in which these galaxies reside. As a result of their large surface densities of star-forming galaxies, the galaxy overdensities identified here are ideal targets for spectroscopic follow-up observations using JWST and/or the Atacama Large Millimeter/sub-millimeter Array (ALMA). Such observations will reveal important details of the galaxy and cluster assembly process, including: the infall of material from the filamentary cosmic web, the interactions between some of the earliest galaxies, the co-evolution of galaxies and their supermassive black holes (SMBHs), the formation of the brightest cluster galaxies (BCGs), the heating and enrichment of the intracluster medium (ICM), and the build-up of the intracluster light (ICL).

%% Start of section five.
\section*{Acknowledgements}
\label{SectionFive}

This work is based on observations made with the NASA/ESA/CSA James Webb Space Telescope. The data were obtained from the Mikulski Archive for Space Telescopes (MAST) at the Space Telescope Science Institute, which is operated by the Association of Universities for Research in Astronomy, Inc., under NASA contract NAS 5-03127 for JWST. These observations are associated with program \#1180, 1181, 1210, 1286, 1895 and 1963. The specific observations analyzed here can be accessed via \dataset[DOI: 10.17909/z2gw-mk31]{https://doi.org/10.17909/z2gw-mk31} \citep{JADES_DR2_DOI} and \dataset[DOI: 10.17909/T91019]{https://doi.org/10.17909/T91019} \citep{HLF_DOI}. The authors sincerely thank the FRESCO team (PI: Pascal Oesch) and the JEMS team (PI: Christina Williams, Sandro Tacchella, and Michael Maseda) for developing and executing their observing programs. We additionally thank the the anonymous referee for providing constructive and insightful suggestions that helped strengthen this paper. This work is also based (in part) on observations made with the NASA/ESA Hubble Space Telescope. The data were obtained from the Space Telescope Science Institute, which is operated by the Association of Universities for Research in Astronomy, Inc., under NASA contract NAS 5–26555. Additionally, this work made use of the {\it lux} supercomputer at UC Santa Cruz which is funded by NSF MRI grant AST 1828315, as well as the High Performance Computing (HPC) resources at the University of Arizona which is funded by the Office of Research Discovery and Innovation (ORDI), Chief Information Officer (CIO), and University Information Technology Services (UITS).

We respectfully acknowledge the University of Arizona is on the land and territories of Indigenous peoples. Today, Arizona is home to 22 federally recognized tribes, with Tucson being home to the Oodham and the Yaqui. Committed to diversity and inclusion, the University strives to build sustainable relationships with sovereign Native Nations and Indigenous communities through education offerings, partnerships, and community service.

FS, CNAW, MJR, GHR, DJE, BR, BDJ, and EE acknowledge support from the JWST/NIRCam contract to the University of Arizona, NAS 5-02015. CW is supported by the National Science Foundation through the Graduate Research Fellowship Program funded by Grant Award No. DGE-1746060. DJE is supported as a Simons Investigator. WB, RM, and JW acknowledge support by the Science and Technology Facilities Council (STFC), as well as the European Research Council (ERC) through Advanced Grant 695671 “QUENCH”. AJB acknowledges funding from the "FirstGalaxies" Advanced Grant from the ERC under the European Union’s Horizon 2020 research and innovation programme (Grant agreement No. 789056). RM and JW acknowledge support by the UKRI Frontier Research grant RISEandFALL. RM acknowledges funding from a research professorship from the Royal Society.

\facilities{HST (ACS), JWST (NIRCam)}

\software{\texttt{AstroPy} \citep[][]{Astropy:2013, Astropy:2018, Astropy:2022}, \texttt{Cloudy} \citep{Ferland:2013, Byler:2017}, \texttt{dynesty} \citep[][]{Speagle:2020}, \texttt{FSPS} \citep[][]{Conroy:2009, Conroy:2010}, \texttt{JWST Calibration Pipeline} \citep{JWST_Pipeline_v1p9p6, JWST_Pipeline_v1p11p2}, \texttt{Matplotlib} \citep[][]{Matplotlib:2007}, \texttt{NumPy} \citep[][]{NumPy:2011, NumPy:2020}, \texttt{Pandas} \citep[][]{Pandas:2022}, \texttt{photutils} \citep[][]{Bradley:2022}, \texttt{Prospector} \citep[][]{Johnson:2021}, \texttt{python-FSPS} \citep[][]{Foreman-Mackey:2014}, \texttt{SciPy} \citep[][]{SciPy:2020}, \texttt{seaborn} \citep[][]{Waskom:2021}, \texttt{TinyTim} \citep[][]{Krist:2011}, \texttt{WebbPSF} \citep[][]{Perrin:2014}}

\bibliographystyle{aasjournal}
\bibliography{main}{}

\end{document}

%% file: figure_set.tex
\figsetstart
\figsetnum{3}

\figsetgrpstart
\figsetgrpnum{3.1}
\figsetplot{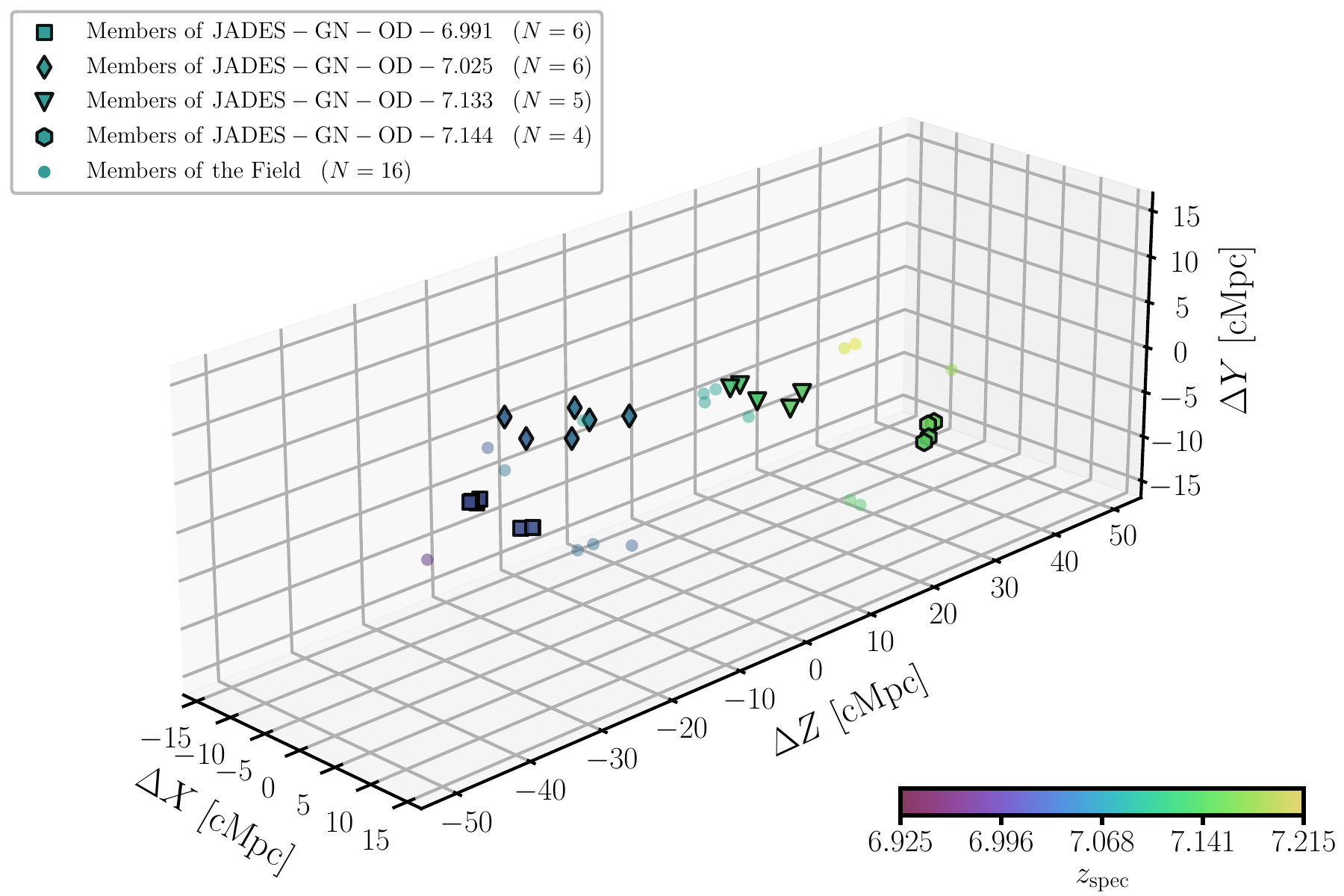}
\figsetgrpnote{3d large-scale structure of the overdense galaxy environment at $6.925 < z_{\mathrm{\,spec}} < 7.215$ in GOODS-N. The on-sky distribution in physical units for the $N = 37$ galaxies that fall within the given comoving volume are color-coded by their spectroscopic redshift. Spatial offsets are measured relative to the median position and redshift of the given sample. The squares represent galaxies that are confirmed members of JADES$-$GN$-$OD$-6.991$, the diamonds represent confirmed members of JADES$-$GN$-$OD$-7.025$, the triangles represent confirmed members of JADES$-$GN$-$OD$-7.133$, the hexagons represent confirmed members of JADES$-$GN$-$OD$-7.144$, and the transparent circles represent confirmed members of the field. These galaxy overdensities reside in a complex environment with connected filamentary structures.}
\figsetgrpend

\figsetgrpstart
\figsetgrpnum{3.2}
\figsetplot{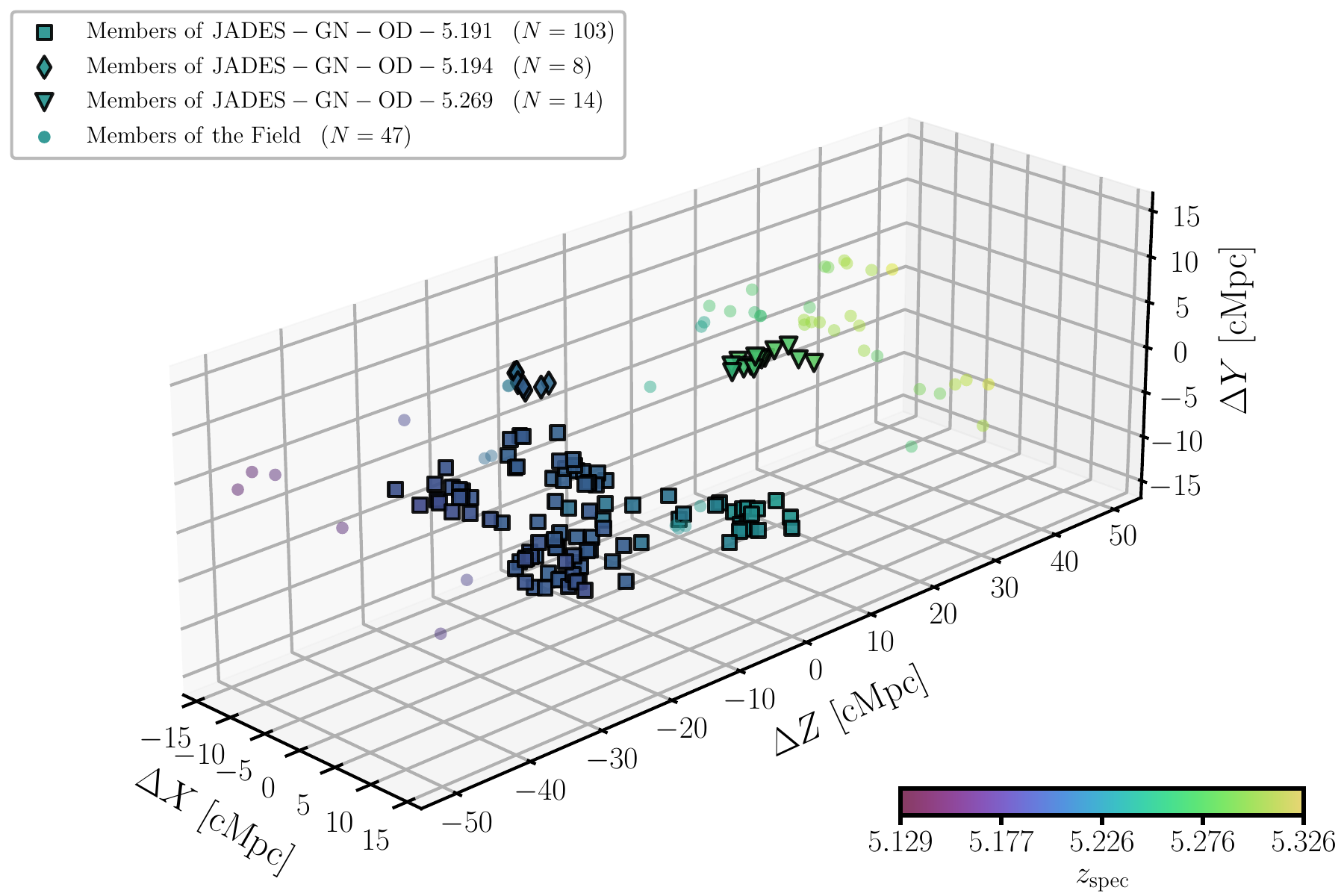}
\figsetgrpnote{3d large-scale structure of the overdense galaxy environment at $5.129 < z_{\mathrm{\,spec}} < 5.326$ in GOODS-N. The on-sky distribution in physical units for the $N = 172$ galaxies that fall within the given comoving volume are color-coded by their spectroscopic redshift. Spatial offsets are measured relative to the median position and redshift of the given sample. The squares represent galaxies that are confirmed members of JADES$-$GN$-$OD$-5.191$, the diamonds represent confirmed members of JADES$-$GN$-$OD$-5.194$, the triangles represent confirmed members of JADES$-$GN$-$OD$-5.269$, and the transparent circles represent confirmed members of the field. These galaxy overdensities reside in a complex environment with connected filamentary structures.}
\figsetgrpend

\figsetgrpstart
\figsetgrpnum{3.3}
\figsetplot{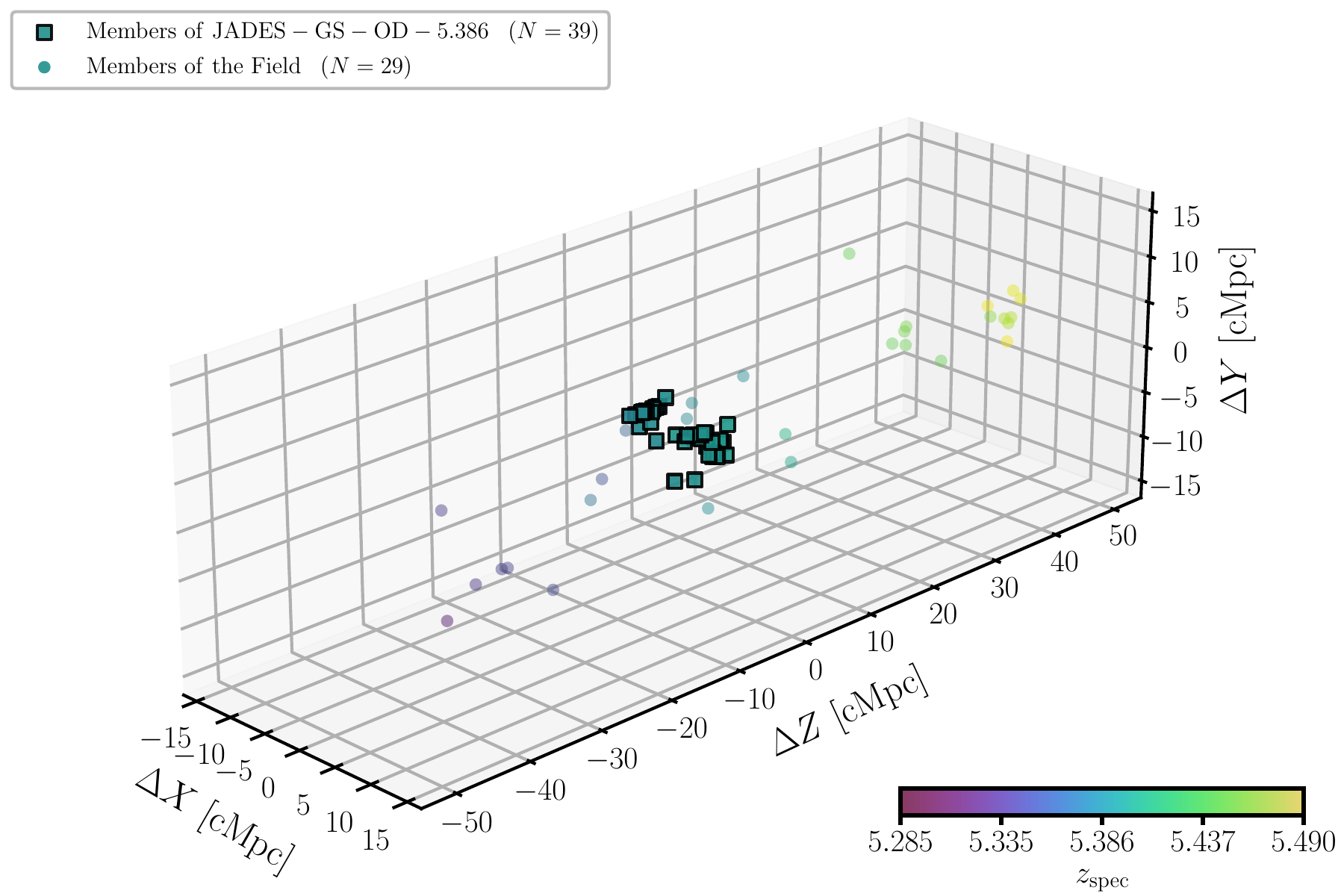}
\figsetgrpnote{3d large-scale structure of the overdense galaxy environment at $5.285 < z_{\mathrm{\,spec}} < 5.490$ in GOODS-S. The on-sky distribution in physical units for the $N = 68$ galaxies that fall within the given comoving volume are color-coded by their spectroscopic redshift. Spatial offsets are measured relative to the median position and redshift of the given sample. The squares represent galaxies that are confirmed members of JADES$-$GS$-$OD$-5.386$ and the transparent circles represent confirmed members of the field. This galaxy overdensity resides in a complex environment with connected filamentary structures.}
\figsetgrpend

\figsetgrpstart
\figsetgrpnum{3.4}
\figsetplot{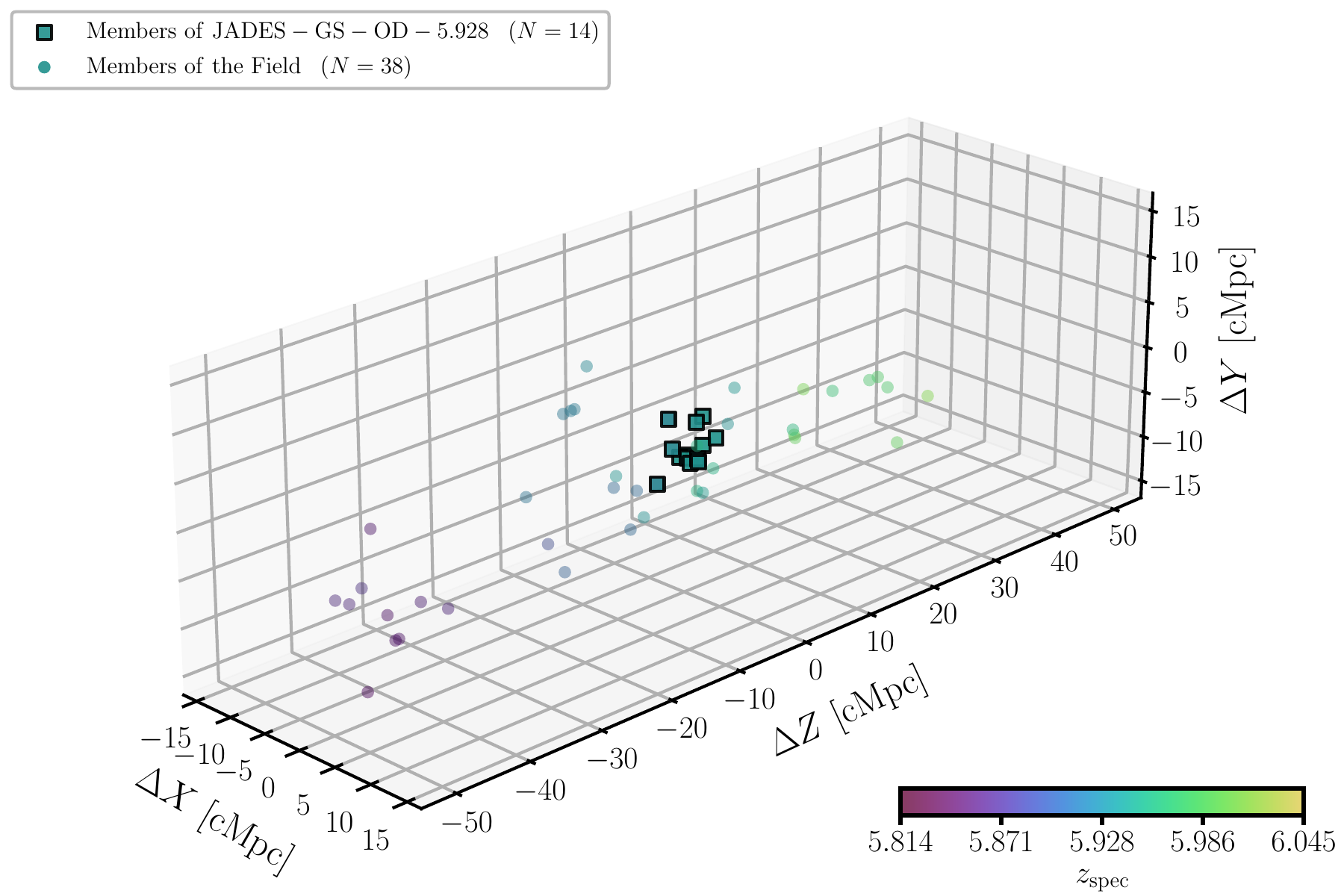}
\figsetgrpnote{3d large-scale structure of the overdense galaxy environment at $5.814 < z_{\mathrm{\,spec}} < 6.045$ in GOODS-S. The on-sky distribution in physical units for the $N = 52$ galaxies that fall within the given comoving volume are color-coded by their spectroscopic redshift. Spatial offsets are measured relative to the median position and redshift of the given sample. The squares represent galaxies that are confirmed members of JADES$-$GS$-$OD$-5.928$ and the transparent circles represent confirmed members of the field.}
\figsetgrpend

\figsetgrpstart
\figsetgrpnum{3.5}
\figsetplot{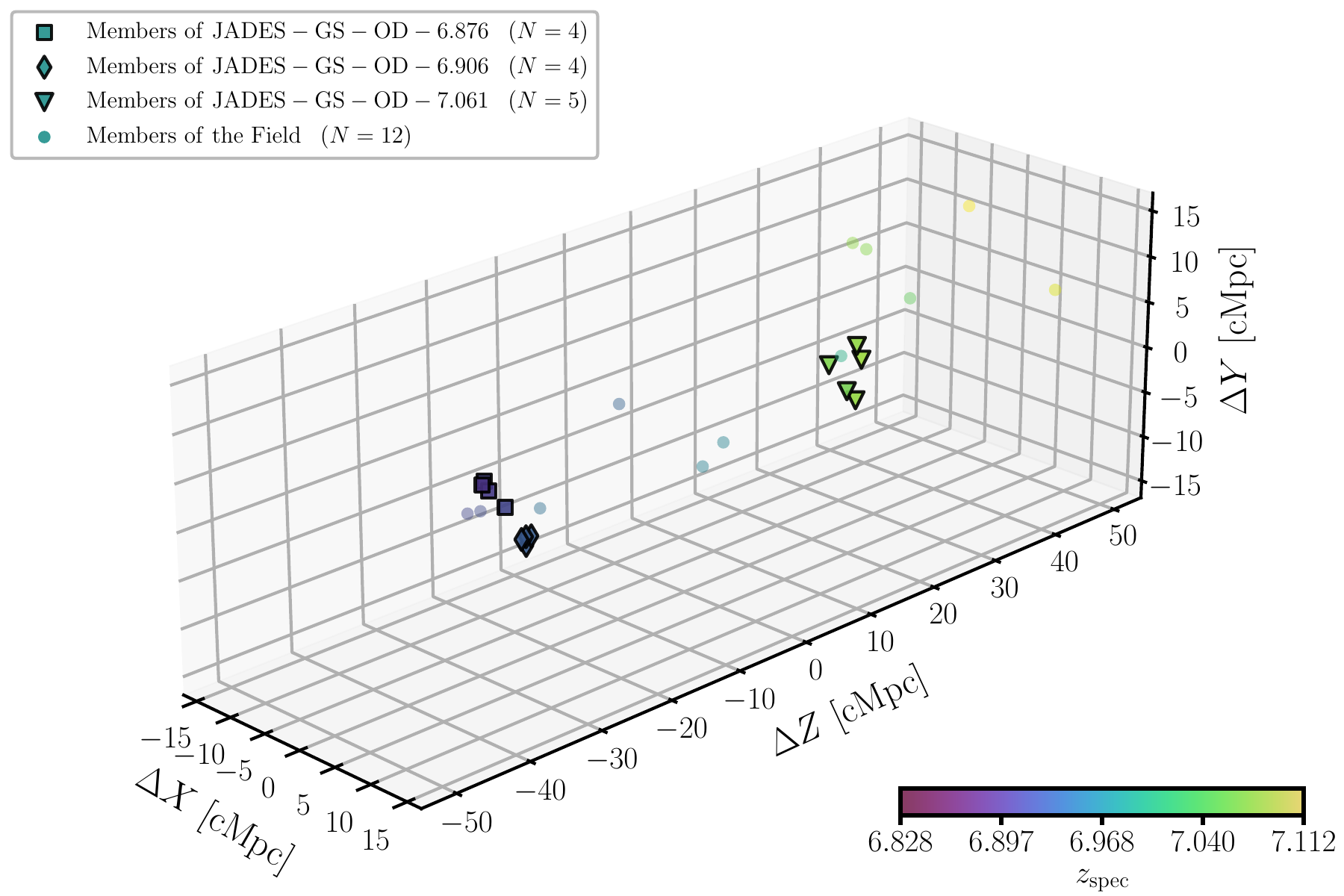}
\figsetgrpnote{3d large-scale structure of the overdense galaxy environment at $6.828 < z_{\mathrm{\,spec}} < 7.112$ in GOODS-S. The on-sky distribution in physical units for the $N = 25$ galaxies that fall within the given comoving volume are color-coded by their spectroscopic redshift. Spatial offsets are measured relative to the median position and redshift of the given sample. The squares represent galaxies that are confirmed members of JADES$-$GS$-$OD$-6.876$, the diamonds represent confirmed members of JADES$-$GS$-$OD$-6.906$, the triangles represent confirmed members of JADES$-$GS$-$OD$-7.061$, and the transparent circles represent confirmed members of the field.}
\figsetgrpend

\figsetgrpstart
\figsetgrpnum{3.6}
\figsetplot{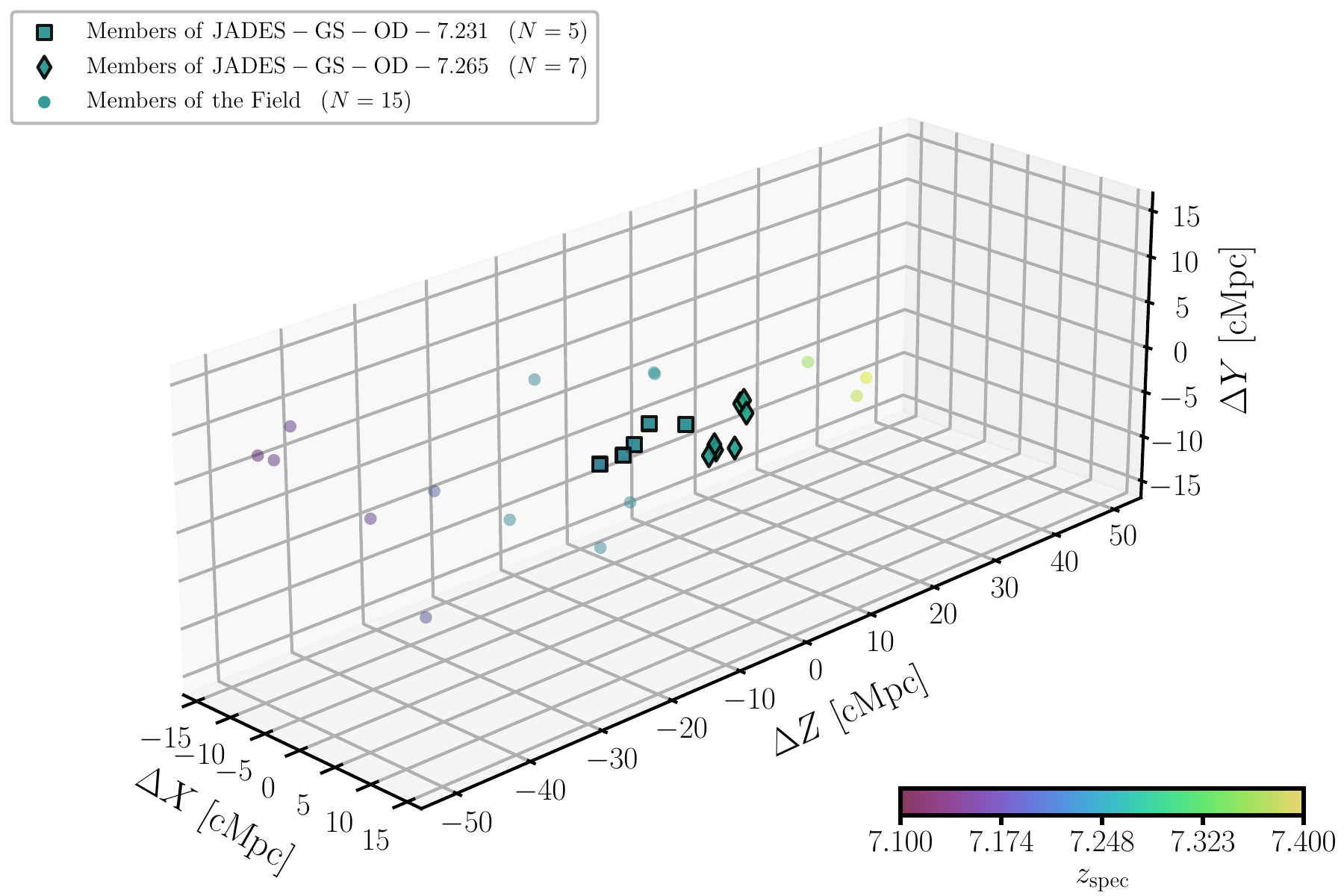}
\figsetgrpnote{3d large-scale structure of the overdense galaxy environment at $7.100 < z_{\mathrm{\,spec}} < 7.400$ in GOODS-S. The on-sky distribution in physical units for the $N = 27$ galaxies that fall within the given comoving volume are color-coded by their spectroscopic redshift. Spatial offsets are measured relative to the median position and redshift of the given sample. The squares represent galaxies that are confirmed members of JADES$-$GS$-$OD$-7.231$, the diamonds represent confirmed members of JADES$-$GS$-$OD$-7.265$, and the transparent circles represent confirmed members of the field. These galaxy overdensities reside in a complex environment with connected filamentary structures.}
\figsetgrpend

\figsetgrpstart
\figsetgrpnum{3.7}
\figsetplot{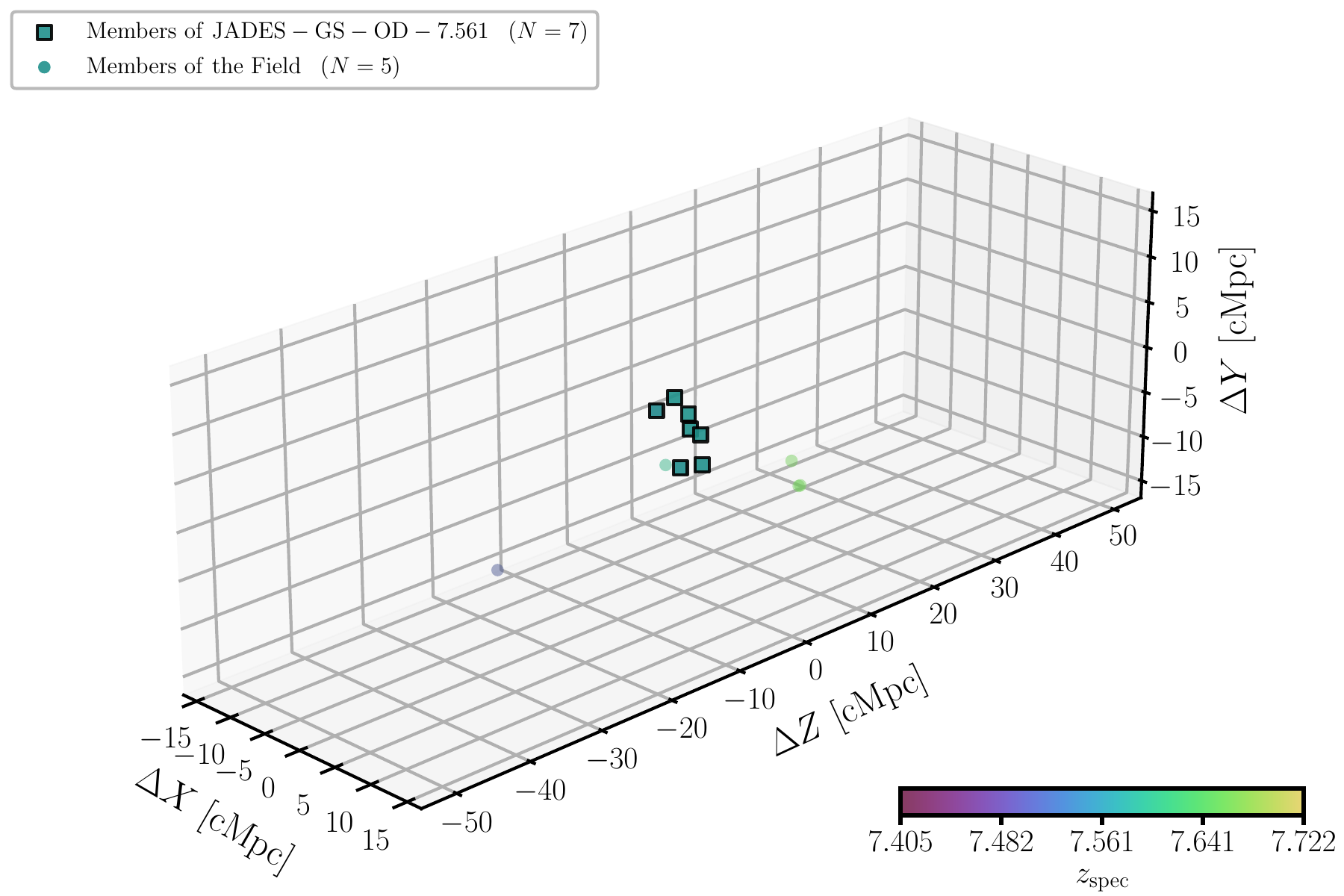}
\figsetgrpnote{3d large-scale structure of the overdense galaxy environment at $7.405 < z_{\mathrm{\,spec}} < 7.722$ in GOODS-S. The on-sky distribution in physical units for the $N = 12$ galaxies that fall within the given comoving volume are color-coded by their spectroscopic redshift. Spatial offsets are measured relative to the median position and redshift of the given sample. The squares represent galaxies that are confirmed members of JADES$-$GS$-$OD$-7.561$ and the transparent circles represent confirmed members of the field.}
\figsetgrpend

\figsetgrpstart
\figsetgrpnum{3.8}
\figsetplot{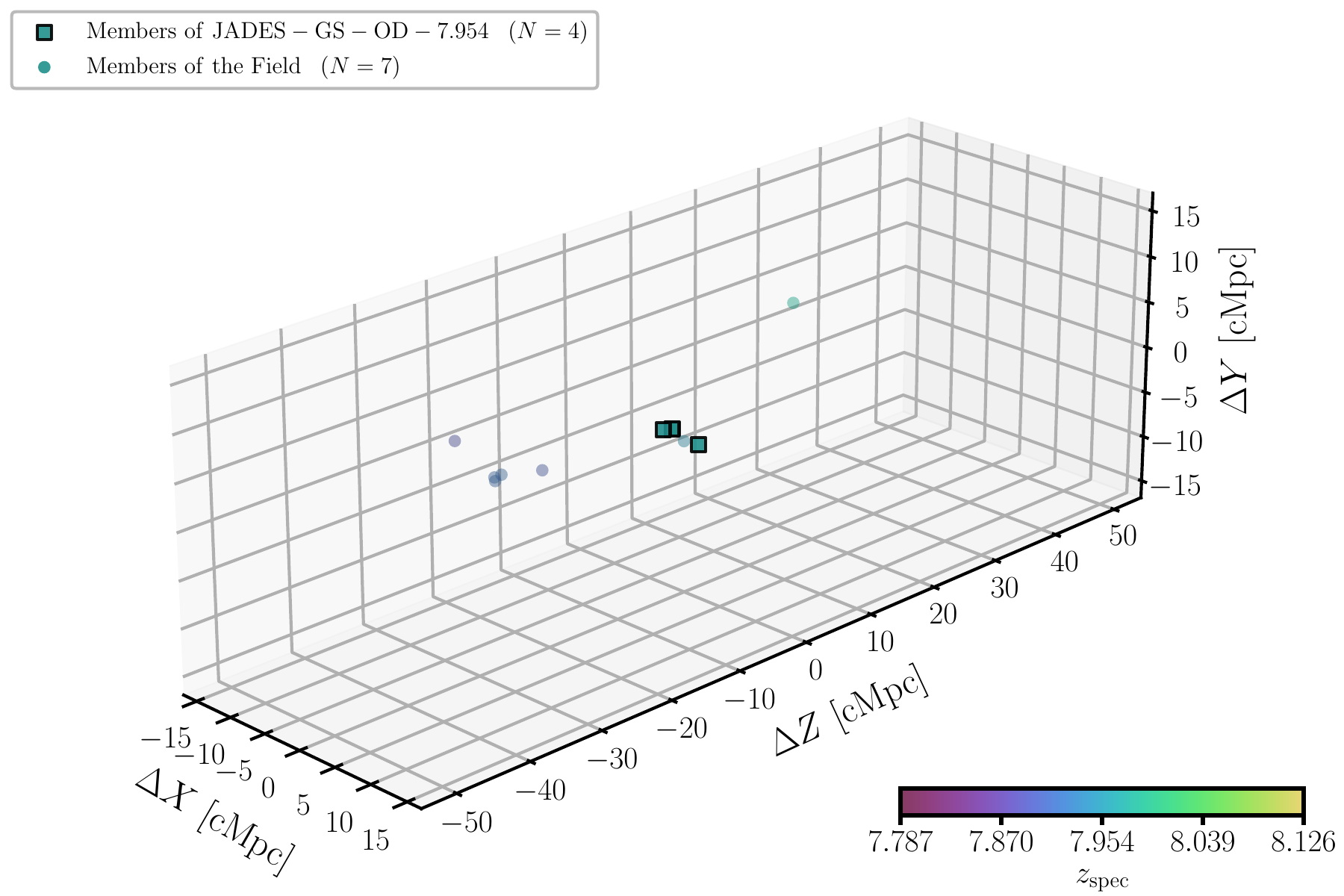}
\figsetgrpnote{3d large-scale structure of the overdense galaxy environment at $7.787 < z_{\mathrm{\,spec}} < 8.126$ in GOODS-S. The on-sky distribution in physical units for the $N = 11$ galaxies that fall within the given comoving volume are color-coded by their spectroscopic redshift. Spatial offsets are measured relative to the median position and redshift of the given sample. The squares represent galaxies that are confirmed members of JADES$-$GS$-$OD$-7.954$ and the transparent circles represent confirmed members of the field.}
\figsetgrpend

\figsetgrpstart
\figsetgrpnum{3.9}
\figsetplot{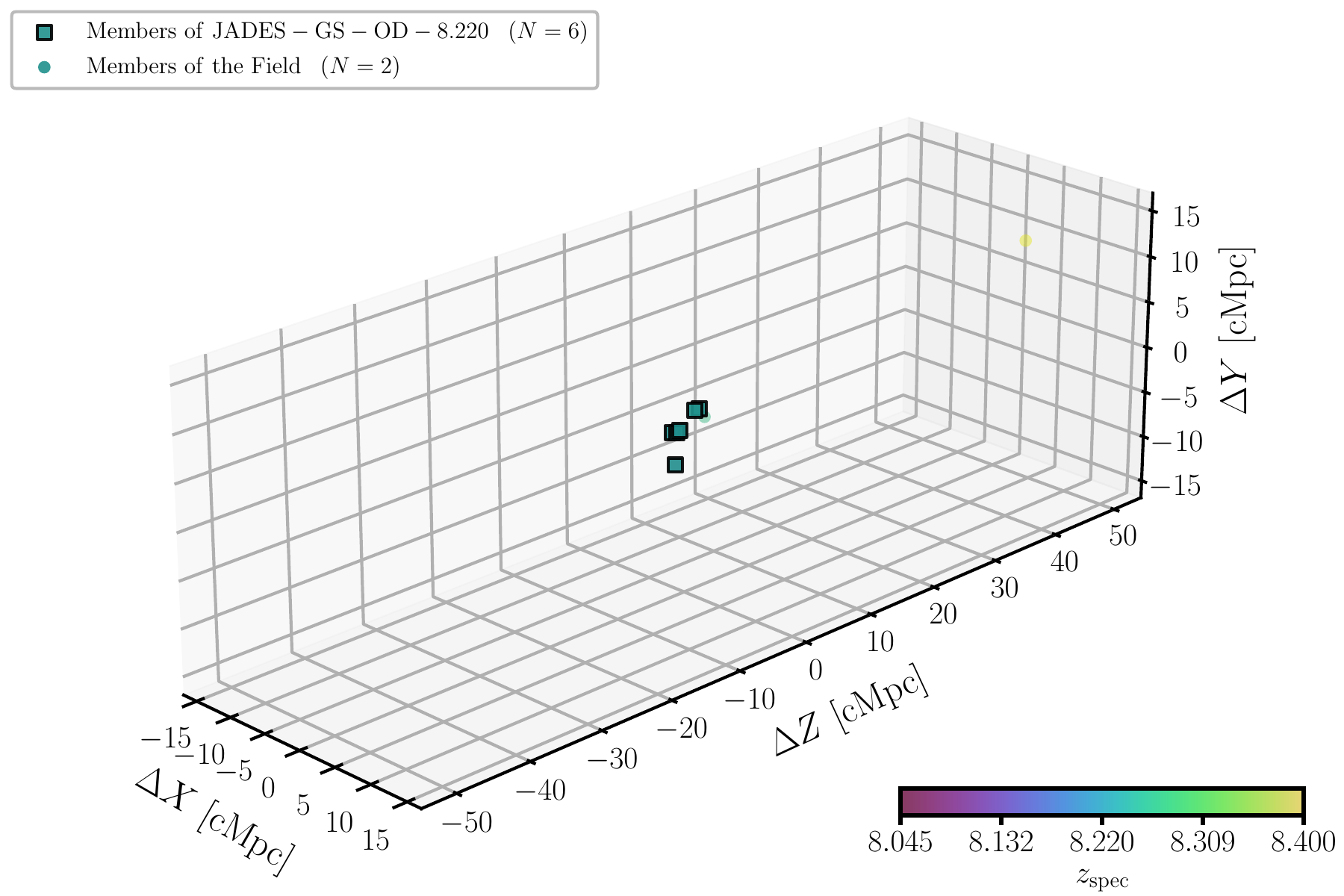}
\figsetgrpnote{3d large-scale structure of the overdense galaxy environment at $8.045 < z_{\mathrm{\,spec}} < 8.400$ in GOODS-S. The on-sky distribution in physical units for the $N = 8$ galaxies that fall within the given comoving volume are color-coded by their spectroscopic redshift. Spatial offsets are measured relative to the median position and redshift of the given sample. The squares represent galaxies that are confirmed members of JADES$-$GS$-$OD$-8.220$ and the transparent circles represent confirmed members of the field. The highest redshift spectroscopically confirmed galaxy overdensity (or protocluster candidate) to date.}
\figsetgrpend

\figsetend

% \begin{figure*}
%     \figurenum{3}
%     \begin{interactive}{animation}{animations/3d_JADES-GN-1.mp4}
%     \centering
%     \includegraphics[width=0.9\linewidth]{figures/3d_JADES-GN-1.pdf}
%     \end{interactive}
%     \caption{3d large-scale structure of the overdense galaxy environment at $6.925 < z_{\mathrm{\,spec}} < 7.215$ in GOODS-N. The on-sky distribution in physical units for the $N = 37$ galaxies that fall within the given comoving volume are color-coded by their spectroscopic redshift. Spatial offsets are measured relative to the median position and redshift of the given sample. The squares represent galaxies that are confirmed members of JADES$-$GN$-$OD$-6.991$, the diamonds represent confirmed members of JADES$-$GN$-$OD$-7.025$, the triangles represent confirmed members of JADES$-$GN$-$OD$-7.133$, the hexagons represent confirmed members of JADES$-$GN$-$OD$-7.144$, and the transparent circles represent confirmed members of the field. These galaxy overdensities reside in a complex environment with connected filamentary structures. \textcolor{red}{The complete figure set (nine animations) is available in the online journal for the rest of the final sample of galaxy overdensities described in Section~\ref{SectionThreeThree}.} \label{figset:3d_JADES-GN-1}}
% \end{figure*}

\begin{figure*}
    \centering
    \figurenum{3}
    \includegraphics[width=1.0\linewidth]{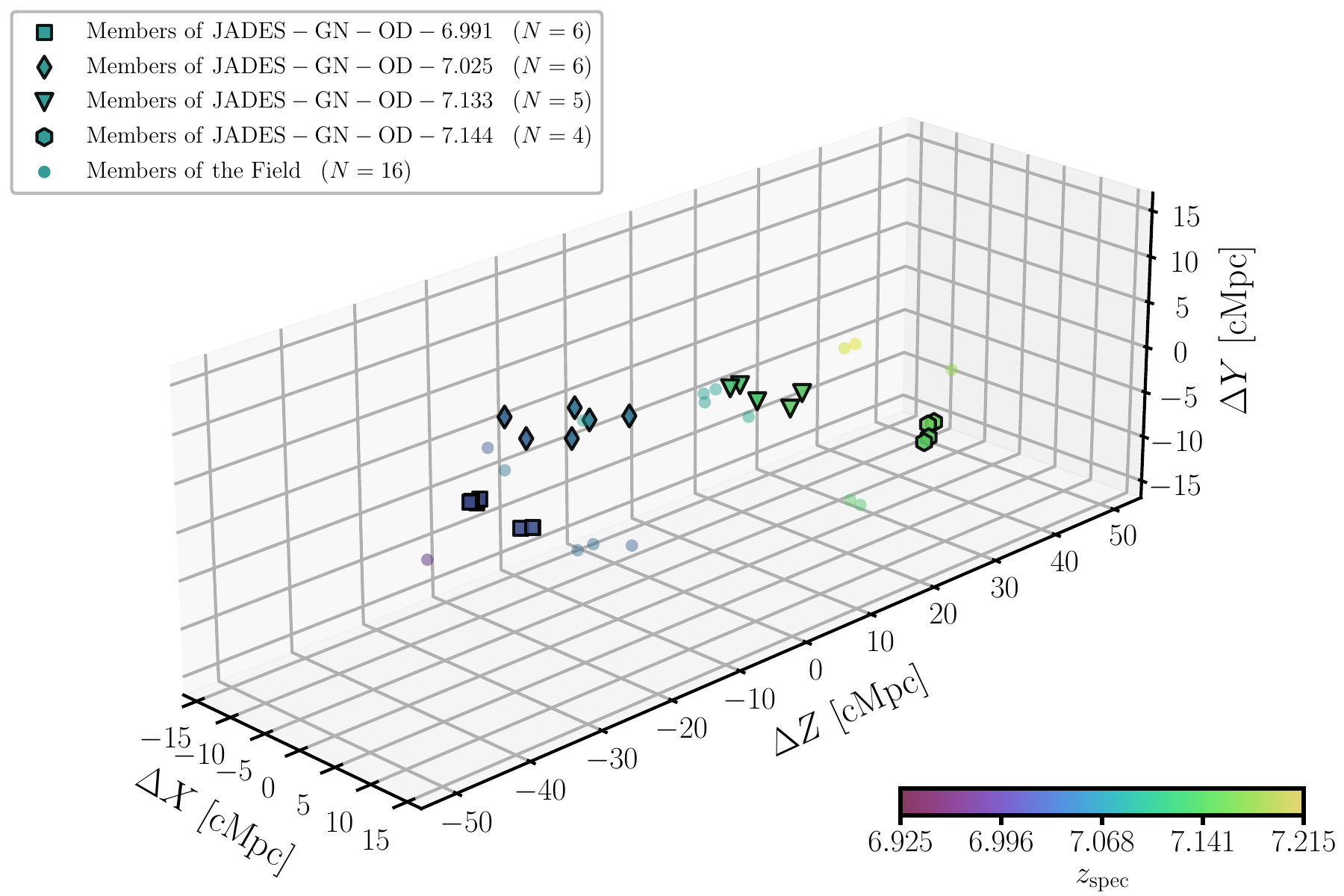}
    \caption{3d large-scale structure of the overdense galaxy environment at $6.925 < z_{\mathrm{\,spec}} < 7.215$ in GOODS-N. The on-sky distribution in physical units for the $N = 37$ galaxies that fall within the given comoving volume are color-coded by their spectroscopic redshift. Spatial offsets are measured relative to the median position and redshift of the given sample. The squares represent galaxies that are confirmed members of JADES$-$GN$-$OD$-6.991$, the diamonds represent confirmed members of JADES$-$GN$-$OD$-7.025$, the triangles represent confirmed members of JADES$-$GN$-$OD$-7.133$, the hexagons represent confirmed members of JADES$-$GN$-$OD$-7.144$, and the transparent circles represent confirmed members of the field. These galaxy overdensities reside in a complex environment with connected filamentary structures. The complete figure set (nine animations) is available in the online journal for the rest of the final sample of galaxy overdensities described in Section~\ref{SectionThreeThree}. \label{figset:3d_JADES-GN-1}}
\end{figure*}